\documentstyle[12pt]{article}
\def \qed{\hfill $\vrule height 2.5mm  width 2.5mm depth 0mm $}
\def\ds{\displaystyle}
\def\mod{\rm mod~}
\def\tr{\rm tr~}
\def\wh{\widehat}
\def\wt{\widetilde} 

\newtheorem{th}{Theorem}

\begin{document}
\title{The  valence bond solid in quasicrystals}
\author{{\Large {Anatol N. Kirillov and Vladimir E. Korepin }} \\
  \\
{\small {\it Steklov Mathematical Institute,}} \\
{\small {\it Fontanka 27, St.Petersburg, 191011, Russia}}}
\date{September 6, 1989}

\maketitle 
\begin{abstract}
A generalized model of Heisenberg quantum antiferromagnet on an arbitrary 
graph is constructed so that the VBS is the unique ground state. The norm 
of the base state and equal time multi point correlation functions are 
computed in terms of generalized hyper geometric functions. For the 
one-dimensional periodic Heisenberg model we present a method of 
computing multi point correlation functions based on the study of a 
commuting family of transfer matrices. The connection of multi point 
correlators with Young tableaux and Gegenbauer polynomials is found.
\end{abstract}
\vskip 0.5cm

{\bf Introduction.}
\vskip 0.5cm

Theory of antiferromagnetism is very important.
We consider the models with Valence Bond Solid ground state [2].
 The study of generalized Heisenberg 
antiferromagnets (see  [3]--[9]) is of great interest .
 A distinguishing
 feature of these models is that their 
Hamiltonians can be represented in the form of a linear combination of 
projections, which makes it possible to explicitly construct the base 
states for the models in question. The models we study are a 
generalization and modification of those considered in [2], [5], [6], 
[7], [8]. They have a valence bound state (abbreviated VBS) as the ground 
state (which is distinct from the Ne\'el ground state). In the first part 
of the work -- \S\S 1--7 -- we devote our attention mainly to questions 
of uniqueness of the VBS. We prove that the VBS is the unique ground 
state of the model in any dimension for a periodic lattice with any 
coordination number. Thus, the N\'eel state is never a ground state for 
the model in question. We have succeeded in reformulating the model on an 
unclosed chain (by introducing special boundary conditions) in such a way 
that the VBS is, as before, the unique ground state. We can construct a 
generalized model of a quantum antiferromagnet (and boundary conditions) 
on an arbitrary graph, so that the VBS is the unique ground state.

The second part of the paper --  \S\S 8--13 --is devoted to the 
computation of the square of the norm of the VBS of the wave function and 
of multi point correlators for the $O(m)$--Heisenberg model on an 
arbitrary graph. The computations make essential use of the properties of 
harmonic polynomials $\Lambda^{A_n}({\bf n})$ (Theorem \ref{t12.1}) 
introduced in \S 9, which are a natural generalization of Gegenbauer 
polynomials to the case of several variables. Theorem B in Appendix B 
plays a central role; it makes it possible to derive recurrence relations 
for the generating function of the correlators (Theorem \ref{t13.2}). We 
show that the Heisenberg $O(m)$--model on one--dimensional or periodic 
chain are connected with a commuting family of transfer matrices (\S 9), 
which provides an alternative method of computing the correlation 
functions (\S\S 10, 11).

An outline of our paper is as follows. In the first two sections we 
present a general construction of models having a unique ground state. 
The models describe the intersection of quantum spins--distinct spins are 
situated at the vertices of an arbitrary graph $\Gamma$. Further, for the 
models constructed we explicitly give the Hamiltonian and present the 
construction of the ground state. Generally speaking, the ground state is 
not unique. In order to formulate a condition for the VBS, we introduce 
the required notation. We denote by $N_0$ the number of vertices and by 
$N_1$ the number of edges of the graph $\Gamma$. Let $S_l$ be the value 
of the spin situated at the vertex of $\Gamma$ with index $l$. We 
consider the vector $2{\bf S}$ whose components with index $l$ is equal 
to $2{\bf S}$. We denote by $\langle kl\rangle$ the edge of $\Gamma$ 
joining the vertices $k$ and $l$. With each edge $\langle kl\rangle$ of 
$\Gamma$ we associate a number $M_{kl}$ which we henceforth call an 
alternating number. We now consider the vector ${\bf M}$ of dimension 
equal to the number of edges $N_1$ of $\Gamma$, whose components with 
index $\langle kl\rangle$ is equal to $M_{kl}$. The (edge--vertex) 
incidence matrix $\wh I$ [10] is an important geometric characteristic of 
$\Gamma$. This is a rectangular matrix of dimension $N_0\times N_1$ whose 
element with index ($n, \langle kl\rangle$) is equal to 
$\delta_{n,k}+\delta_{n,l}$. In the second section we prove the following 
theorem.

{\it Solvability in nonnegative integers $M_{kl}$ (for fixed spins $s_l$) 
of the system of linear equations
\begin{equation}
2{\bf S}={\wh I}\cdot{\bf M} \label{0.1}
\end{equation}
is a necessary and sufficient condition for uniqueness of the ground 
state of generalized Heisenberg magnet corresponding to spins $s_l$ and 
alternating numbers $M_{kl}$.}

In \S\S 3, 5--8 we study conditions for solvability of the system 
(\ref{0.1}) for a fixed collection of spins $s_l$. This problem naturally 
decomposes into two problems: $i)$ solvability of the system in integers 
(the $M_{kl}$ must be integers); $ii)$ solvability of the system in 
positive integers $M_{kl}$. We give a complete solution of the first 
problem. The answer depends in an essential way on whether the graph 
$\Gamma$ is bipartite [10]. We recall that a bipartite graph is 
characterized by the fact that the set of its vertices can be decomposed 
into two subsets $A$ and $B$ so that any edge in $\Gamma$ joins only 
vertices of different subsets. In \S 7 we prove the following theorem.

{\it For a connected, bipartite graph $\Gamma$ the system of equations 
(\ref{0.1}) is solvable in integers $M_{kl}$ if and only if the following 
relation is satisfied:
\begin{equation}
\sum_{a\in A}s_a=\sum_{b\in B}s_b. \label{0.2}
\end{equation}

For a non bipartite graph $\Gamma$ in \S 8 we prove that the system 
(\ref{0.1}) is solvable in integers $M_{kl}$ if and only if the following 
condition is satisfied:}
\begin{equation}
\sum_{l\in\Gamma} s_l\in{\bf Z}. \label{0.3}
\end{equation}

To prove the theorems formulated above it is useful to first consider 
cases where the graph $\Gamma$ is a tree (\S\S 3, 5) or a cycle (\S 6). 
For such graphs we solve the system (\ref{0.1}) explicitly. The result 
obtained are used in the proof of the theorems in the general case.

As concerns solvability of the system (\ref{0.1}) in nonnegative 
integers, we present only necessary conditions (\S 6). The question of 
sufficiency of these conditions remains open.

We consider the model of an antiferromagnet in a quasicrystal separately 
(\S 4). The fact of the matter is that the thermodynamic limit for models 
in a quasicrystal is analogous to the thermodynamic limit in crystal. For 
a one--dimensional quasicrystal we explicitly compute the multi point 
correlation functions.
In the second part of our work we study multi point correlation functions 
for the Heisenberg model on graph $\Gamma$. We begin by considering the 
Heisenberg model on one--dimensional lattice. In this simplest example 
all spins at the nodes of the lattice are equal to 1. The Hamiltonian of 
the model is [5]
\begin{equation}
H=\frac{1}{2}\sum_l\left\{{\wh{\bf S}}(l)\cdot{\wh{\bf S}}(l+1)+
\frac{1}{3}({\wh{\bf S}}(l)\cdot{\wh{\bf S}}(l+1))^2+\frac{2}{3}\right\}. 
\label{0.4}
\end{equation}
Here the ${\wh{\bf S}}(l)$ are the quantum spins (a representation of the 
algebra $0(3)$; $l$ is the index of a node of the lattice). The ground 
state of the model -- the valence bond state -- is described in \S 1. The 
equal time two--point correlator is computed in [2] (in the thermodynamic 
limit):
\begin{equation}
\langle{\wh S}^{a-1}(l_1)\cdot{\wh S}^{a_2}(l_2)\rangle =
\frac{4}{3}\delta^{a_1}_{a_2}(-3)^{-l|l_1-l_2|}. \label{0.5}
\end{equation}
Here the $\wh S^a$ are the components of the vector $\wh{\bf S}$ 
($a=1,2,3$). We prove that for the model under consideration the equal time 
multi point correlator reduces to the product of the two--point 
correlators (\ref{0.5}). More precisely, we have the following:

1) If the number of spins is odd, then
\begin{equation}
\left\langle\prod_{j=1}^N\wh S^{a_j}(l_j)\right\rangle =0, \ \ 
N\equiv 1~(\mod 2). \label{0.6}
\end{equation}
We suppose here and below that all the coordinates $l_j$ are distinct.

2) If the number of spins $N$ is even, $l_N>l_{N-2}>\cdots >l_2>l_1$, 
then we prove that
\begin{equation}
\left\langle\prod_{j=1}^N\wh S^{a_j}(l_j)\right\rangle =\prod_{r=1}^{N/2}
\langle\wh S^{a_{2r}}(l_{2r})\cdot \wh S^{a_{2r-1}}(l_{2r-1})\rangle . 
\label{0.7}
\end{equation}

A generalization of the model (\ref{0.4}) to the case of higher spin $S$ 
is given in [5], [8]. In this case the Hamiltonian of the model is given 
by formulas (\ref{1.6}) and (\ref{1.8}) of the text. The equal time 
two--point correlator (in the thermodynamic limit) for the Heisenberg 
model of higher spin $S$ is computed in [5]:
\begin{equation}
\langle\wh S^{a_1}(l_1)\wh S^{a_2}(l_2)\rangle =\frac{(s+1)^2}{3}
\delta_{a_2}^{a_1}\left(\frac{-s}{s+2}\right)^{|l_1-l_2|}. \label{0.8}
\end{equation}

In the present paper we compute the equal time multi point correlator (in 
the thermodynamic limit) for a model of spin $S$. The precise formulation 
is presented in \S 11 and differs basically from (\ref{0.7}). As an 
example, we present a formula for the four--point correlator
\vskip 0.4cm

$\langle\wh S^{a_1}(l_1)\wh S^{a_2}(l_2)\wh S^{a_3}(l_3)
\wh S^{a_4}(l_4)\rangle=$
\begin{eqnarray}
&=&\frac{(s+1)^4}{9}\delta^{a_1}_{a_2}\delta^{a_3}_{a_4}
\left(\frac{-s}{s+2}\right)^{l_2-l_1+l_4-l_3}\nonumber \\
&+&\frac{(s+1)^4}{15}\left(\frac{-s}{s+2}\right)^{l_2-l_1+l_4-l_3}
\left(\frac{s(s-1)}{(s+2)(s+3)}\right)^{l_3-l_2} \label{0.9}\\
&\times &\left\{\delta^{a_1}_{a_4}\delta^{a_2}_{a_3}+
\delta^{a_1}_{a_3}\delta^{a_2}_{a_4}-\frac{2}{3}\delta^{a_1}_{a_2}
\delta_{a_4}^{a_3}\right\}.\nonumber
\end{eqnarray}
Here $l_4>l_3>l_2>l_1$.

We note that in the case of a one--dimensional, periodic chain the models 
considered (\ref{0.4}) and (\ref{1.6})--(\ref{1.8}) have a unique ground 
state. For a finite open chine the ground state is no longer unique. In 
the first part of the present paper we analyze conditions under which the 
models considered have a unique ground state. For example, for the model 
(\ref{1.6})--(\ref{1.8}) on an open chain the two boundary spins $s$ must 
be replaced by spins equal to $s/2$. These modified models have a unique 
ground state. such modification does not affect the thermodynamic limit, 
but it simplifies the computation of the correlation functions for finite 
chains.

In the first part of the paper we prove that there exists a natural, 
inhomogeneous generalization of the models considered above (distinct 
spins are situated at the vertices of an arbitrary graph) with 
presentation of uniqueness of the ground state. One--dimensional 
quasicrystals are examples of such models. We also compute the 
multi point, equal time correlators (in the thermodynamic limit) for an 
inhomogeneous model on a chain (see \S 11). In [2] it was shown that the 
quantum models considered above are equivalent to a one--dimensional, 
classical, modified Heisenberg model, which is useful in computing the 
correlators. We give a description of the modified model in \S 8. Using 
the integral operators $\wh K_M$ (see \S 9), we construct a transfer 
matrix for the classical model and find the spectrum and eigenfunction 
of the operators $\wh K_M$ in \S 9. This enables us to compute the norm 
of the wave function (see \S 10) and also to find the multi point 
correlators (\S 11). In the following \S 12 we give a generalization to 
the case of the group $O(m)$ of the modified Heisenberg model considered 
earlier. In \S 13 we consider the quantum and corresponding classical 
Heisenberg model for a multidimensional lattice. The inhomogeneous 
Heisenberg model (arbitrary spins are situated at the vertices) for a 
complete graph (all the vertices of the graph are joined by edges) is a 
natural algebraic object. We consider the generating function for the 
correlators and derive for it recurrence relations which make it possible 
in principle to find it for the complete graph (and hence for an 
arbitrary graph). We show that the generating function of the correlators 
can be expressed in terms of generalizes hyper geometric functions.

We consider in more detail a one--dimensional quasicrystal [12]. There 
exists a quasi periodic covering by a direct infinite sequence of two 
intervals (short and long). One of the examples of such a covering is 
connected with the golden section $\tau =(1+\sqrt 5)/2$. The position of 
the end of the $l$th segment of the covering is found by the formula
\begin{equation}
x_l=l+\frac{1}{\tau }\left[\frac{l}{\tau }+\alpha\right]. \label{0.10}
\end{equation}
Here $\alpha$ is real parameter. We denote by $p_L$ ($p_s$) the 
probability of the occurrence of long (short) interval [12]. We construct 
a Heisenberg antiferromagnet for the quasi periodic lattice (\ref{0.10}). 
To each long segment we assign an alternating number $M_L$ and to each 
short segment we assign a number $M_s$. The spins (\ref{2.2}) can then 
assume only the three values
\begin{equation}
s(l)=\left\{M_L,M_S,(M_L+M_S)/2\right\} \label{0.11}
\end{equation}
depending on the position of the vertex $l$. The Hamiltonian of the model 
is given by formulas (\ref{2.3}), (\ref{2.4}), while the ground VBS state 
is given by formula (\ref{2.5}). In \S 10 we explicitly compute the 
correlation functions -- for a finite chain and in the thermodynamic 
limit. We present the answer for the asymptotic of the two--point 
correlator of a quasicrystal:
\begin{equation}
\langle\wh S^{a_1}(l_1)\wh S^{a_2}(l_2)\rangle =\frac{1}{3}
\delta^{a_1}_{a_2}(s(l_1)+1)(-1)^{l_2-l_1}\exp\left\{-m(l_2-l_1)\right\}. 
\label{0.12}
\end{equation}
Here $(l_2-l_1)\to\infty$,
\begin{equation}
m=p_s\log\left(1+\frac{2}{M_S}\right)+p_L\log\left( 1+\frac{2}{M_L}\right).
\label{0.13}
\end{equation}

We are deeply grateful to L.D.~Faddeev and N.Leskova
 for his constant interest in the 
work and for many useful remarks.The first version of this paper was
published as a preprint in 1988 [1].

\vskip 0.5cm
{\bf \S 1. A generalized model of an antiferromagnet. Uniqueness of the 
 ground state.}
\vskip 0.5cm

 We consider a periodic lattice in $D$--dimensional Euclidean space. The 
 interacting quantum spins $\bf{\wh S_j}$ are situated at the nodes of the 
 lattice ($l$ is the index of a node). At each node the spins are a 
 representation of the algebra $SU(2)$:
 \begin{equation}
 [\wh S_l^a,\wh S_k^b]=i\delta_{lk}\epsilon^{abc}\wh S_l^c. \label{1.1}
 \end{equation}
 Here the spin index $a$ assumes the values $a=1,2,3$. The lower index 
 $l$ numbers the nodes of the lattice. We express the spins in terms of 
 two independent, canonical Bose fields on the lattice -- $a_l$ and 
 $b_l$. Their commutation relations are standard:
 \begin{equation}
 [a_l,a_k^+]=\delta_k^l, \ [b_l,b_k^+]=\delta_k^l, \ [a_l,b_k]=0. 
\label{1.2}
 \end{equation}
 The components of the spin can be expressed as follows:
 $$\wh S_l^+=a_l^+b_l, \ \wh S_l^-=b_l^+a_l,
 $$
 \begin{equation}
 \wt S_l^3=\frac{1}{2}(a_l^+a_l-b_l^+b_l).\label{1.3}
 \end{equation}
 The value of the spin at the node with index $l$ is an eigenvalue of the 
 operator
 \begin{equation}
 {\bf\wh S}_l=\frac{1}{2}(a_l^+b_l+b_l^+b_l). \label{1.4}
 \end{equation}
 (In a quasicrystal different spins are present at different nodes.)
 
 Thus, the spin operators (\ref{1.3}) act in Fock space whose vectors have 
 the form
 \begin{equation}
 P(a_l^+,b_l^+)~|~0\rangle . \label{1.5}
 \end{equation}
 Here $P$ is a polynomial in the variables $a_l^+$ and $b_l^+$ ($l$ runs 
 through all nodes of the lattice). We seek eigenfunction of the 
 operator ${\bf\wh S}_l$:
 $${\bf\wh S}_lP(a^+,b^+)~|~0\rangle =s_lP(a^+,b^+)~|~0\rangle .
 $$
 
 This relation means that the polynomial $P$ is a homogeneous function of 
 the variables $a^+_l$ and $b^+_l$ (for given $l$) of degree $2s_l$, 
 i.e., the polynomial $P$ can be represented in the form
 $$P=\sum_{k=0}^{2s_l}(a^+_l)^k(b^+_l)^{2s_j-k}\wt P_k.
 $$
 Here the polynomial $\wt P_k$ does not depend on the variables at the 
 $l$th node.
 
 The Hamiltonian of the model describes the interaction only of spins 
 situated at the nearest nodes of the lattice (we denote this by $\langle 
 kl\rangle$ ):
 \begin{equation}
 H=\sum_{\langle kl\rangle}H(k,l). \label{1.6}
 \end{equation}
 The Hamiltonian density $H(k,l)$ is usually described in terms of powers 
 of the scalar product of spins $({\bf S}_k\cdot{\bf S}_l)^n$.  Instead 
 of this  we use a special basis of polynomials in 
 $({\bf S}_k\cdot{\bf S}_l)$. These are the projections $\Pi_J(k,l)$ 
 onto the state with fixed spin $J$. They can be found from the following 
 system of linear equations:
 \begin{equation}
 ({\bf S}_k\cdot{\bf S}_l)^n=\sum_{J=|s_k-s_l|}^{s_k+s_l}\Pi_J(k,l)
 \left[\frac{1}{2}J(J+1)-\frac{1}{2}s_l(s_l+1)-\frac{1}{2}s_k(s_k+1)
 \right]^n, \label{1.7}
 \end{equation}
 $n=0,1,2,\ldots ,2s_{\min}$.\\
 Here $s_k$ and $s_l$ are the magnitudes of the spins at the nodes $k$ 
 and $l$, while $s_{\min}$ is the least of these two values. The system 
 of equations (\ref{1.7}) can easily be solved for projections 
 $\Pi_J(k,l)$; we find the following expression:
 \begin{equation}
 \Pi_J(k,l)=\prod_{j\ne J, ~|s_k-s_l|\le j\le s_k+s_l}\frac{\wh S^2-j(j+1)}
 {J(J+1)-j(j+1)}. \label{1.8}
 \end{equation}
 Here $\wh S^2=(\wh S_k+\wh S_l)^2$. The projection (\ref{1.8}) is a 
 polynomial in $(S_k,S_l)$ of degree $2s_{\min}$. In order to produce the 
 Hamiltonian of [1], [3], we suppose that the magnitude of spin at each 
 node is the same and equal to $s$ (we give up this assumption in next 
 section). The Hamiltonian density is
 \begin{equation}
 H(k,l)=\sum_{J=2s+1-M}^{2s}A_J\Pi_J(k,l). \label{1.9}
 \end{equation}
 Here $M$ is a positive integer, $1\le M\le 2s$. (We shall see below that 
 this is an important parameter of the theory.) It is important that the 
 following relation be satisfied:
 \begin{equation}
 2s=zM. \label{1.10}
 \end{equation}
 Here $z$ is the coordination number of the node (the number of nearest 
 neighbors). The coefficients $A_J$ are positive real numbers--parameters 
 of the model. The model has thus been determined. we call it the AKLT 
 model [2]. We present examples. For a one--dimensional, periodic lattice 
 $z=2$, and for the least value $s=1$, $M=1$
 \begin{equation}
 H=A\sum_{\langle kl\rangle}\left\{({\bf S}_l\cdot{\bf S}_k)^2_-+3({\bf 
 S}_l\cdot{\bf S}_k)+2\right\} . \label{1.11}
 \end{equation}
 In the two--dimensional case on a hexagonal lattice $z=3$, and for the 
 least value $s=3/2$, $M=1$
 \begin{equation}
 H=A\sum_{\langle kl\rangle}\left\{({\bf S}_l\cdot{\bf 
 S}_k)+\frac{116}{243}({\bf 
 S}_l\cdot{\bf S}_k)^2+\frac{16}{243}({\bf S_l}\cdot{\bf 
 S}_k)^3\right\} . \label{1.12}
 \end{equation}
 
 We now return to the Hamiltonian (\ref{1.9}) and construct the ground 
 state.
\begin{th} \label{t1.1} The Hamiltonian (\ref{1.9}) has a unique ground state of the 
following form:
\begin{equation}
|\psi\rangle ={\rm const}\prod_{\langle kl\rangle}(a^+_kb^+_l-a^+_lb^+_k)^M.
\label{1.13}
\end{equation}
\end{th}
{\it Proof.} It is obvious that $H\ge 0$ and also that $\Pi_J(k,l)\ge 0$. Thus, 
if there exists a solution $|\psi\rangle$ of the equation
\begin{equation}
H|\psi\rangle =0 \label{1.14}
\end{equation}
then it is ground state of the Hamiltonian (\ref{1.9}). Due to positivity 
equation (\ref{1.14}) is equivalent to the set of equations
\begin{equation}
\Pi_J(k,l)|\psi\rangle =0 \label{1.15}
\end{equation}
for any pair of nearest nodes $\langle kl\rangle$ and for any $J$ in the 
interval $2s+1-M\le J\le 2s$. This implies that in adding spins $S_k$ and 
$S_l$ there are no projections onto states with complete spin $J$, where
$2s+1-M\le J\le 2s$. We now use the following theorem, whose proof we 
present in Appendix A.  
\begin{th} \label{t1.2} (on addition of spins). We suppose that after addition of 
spins $\wh S_l$ and $\wh S_k$ there arises a state $|\psi\rangle$ with 
zero projections onto spins $J$ of the interval
\begin{equation}
s_l+s_k+1-M\le J\le s_l+s_k. \label{1.16}
\end{equation}
Then the polynomial $|\psi\rangle$ is divisible by 
$(a_k^+b_l^+-a_l^+b_k^+)^M$.
\end{th}

We continue the proof of Theorem \ref{t1.1}. We recall that we are seeking 
a ground state $|\psi\rangle$ in the form (\ref{1.5}). It follows from 
Theorem \ref{t1.2} that the polynomial $P(a^+,b^+)$ is divisible by 
$(a^+_kb^+_l-a^+_lb^+_k)^M$ for each edge $\langle kl\rangle$. Hence,
\begin{equation}
|\psi\rangle =P(a^+,b^+)|0 \rangle =\left\{\prod_{\langle kl\rangle}
(a^+_kb^+_l-a^+_lb^+_k)^M\right\}\wt P(a^+,b^+)|0 \rangle . \label{1.17}
\end{equation}
Here $\wt P(a^+,b^+)$ is another polynomial. We now compute the magnitude 
of the spin of the state $|\psi\rangle$ at the node with index $m$. This 
is an eigenvalue of the operator
\begin{equation}
\wh S_m=\frac{1}{2}(a^+_ma_m+b^+_mb_m), \label{1.18}
\end{equation}
in other words, $\wh S_m|\psi\rangle =s_m|\psi\rangle$.

>From this it follows that the polynomial $P$ is a homogeneous function of 
the variables $a^+_m$ and $b_m^+$. Applying the operator (\ref{1.18}) and 
(\ref{1.17}), we obtain
\begin{equation}
2s_m=Mz+\Delta_m. \label{1.19}
\end{equation}
Here $\Delta_m\ge 0$ is the degree of homogeneity of the new polynomial 
$\wt P$ in the variables $a^+_m$ and $b^+_m$. Comparing (\ref{1.19}) and 
(\ref{1.10}), we find that $\Delta_m=0$, i.e., the polynomial $\wt P$ 
does not depend on the variables $a^+_m$ and $b^+_m$. Thus, $\wt 
P=$const as asserted. This means that equation (\ref{1.14}) has a unique 
solution
\begin{equation}
|\psi\rangle =\prod_{\langle kl\rangle}(a^+_kb^+_l-a^+_lb^+_k)^M. 
\label{1.20}
\end{equation}
The theorem on the existence and uniqueness of a ground state of the 
Hamiltonian (\ref{1.9}) has thus been proved.

\qed

The wave function (\ref{1.20}) realized the valence bound state. In [1], 
[2], [5], [6] it is called a VBS (valence bond state).

We shall now try to construct a model for an unclosed chain so that the 
uniqueness theorem is preserved. We thus consider a one--dimensional 
chain of $N$ nodes $l=1,2,\ldots ,N$. We observe immediately that for the 
interior nodes ($l=2,\ldots ,N-1$) the coordination number $z=2$, while 
for the boundary nodes ($l=1$, $N$) $z=1$. From the relation $2s_l=z_lM$ 
(for $M=1$) it follows that for interior nodes $s_1=1$ while for the 
boundary nodes $s_N=1/2$.

We take the Hamiltonian in the form
\begin{equation}
H=\sum_{l=1}^{N-1}\Pi_{s_l+s_{l+1}}(l,l+1). \label{1.21}
\end{equation}
We note that for interior edges $\Pi$ is the projection onto spin 2: 
$\Pi_{s_l+s_{l+1}}=\Pi_2$, while for boundary edges it is the projection 
onto spin 3/2 ($\Pi_{s_l+s_{l+1}}=\Pi_{3/2}$ for $l=1$ and $l=N-1$). This 
is the difference between our Hamiltonian and that proposed in [2], [5], 
[6], [7]. In those papers it was assumed that $s_1=s_N=1$, and the ground 
state is fourfold degenerate. For the hamiltonian (\ref{1.21}) it is easy 
to prove that the ground state is unique. In the next section we 
construct a model with a unique ground state on an arbitrary graph.
\vskip 0.5cm

{\bf \S 2. The model on an arbitrary graph.}
\vskip 0.5cm

We shall construct a generalized model of an antiferromagnet on an 
arbitrary graph $\Gamma$. We do this so that the theorem on existence and 
uniqueness of the wave function of ground state (VBS) is preserved. We 
consider an arbitrary graph. It consists of $N_0$ vertices. Some of the 
vertices are connected by edges. We denote the number of edges by $N_1$. 
The edge--vertex incidence matrix $\wh I$ plays an important role below. 
This is matrix with $N_0$ rows and $N_1$ columns. Each of its matrix 
elements is equal to 0 or 1. We define $\wh I$ more precisely. Each row 
of $\wh I$ is connected with specific vertex of the graph $\Gamma$; each 
column is connected with an edge. If a vertex belongs to an edge, then 
the corresponding matrix element is equal to 1; otherwise it is equal to 
0. We have thus defined the incidence matrix. We number all vertices of 
the graph, for example, by the letter $l$. We number edges by pair of 
letters (for example, $\langle kl\rangle$) denoting the vertices 
belonging to the given edge. We begin the construction of a generalized 
model of an antiferromagnet. To each vertex $l$ we assign a spin $s_l$ 
(the integer $2s_l>0$), and to each edge $\langle kl\rangle$ we assign 
another positive integer $M_{kl}>0$. At each node these numbers must be 
connected by relation
\begin{equation}
2s_l=\sum_{\langle kl\rangle}M_{kl}. \label{2.1}
\end{equation}
Here the summation goes over all edges abutting the node $l$. This 
relation can be written with the help of the incidence matrix as
\begin{equation}
2{\bf S}={\wh I}\cdot{\bf M}. \label{2.2}
\end{equation}
Here $\bf S$ is an $N_0$--component vector (its components are equal to 
$s_l$), while $\bf M$ is an $N_1$--component vector (its components are 
equal to $M_{kl}$). We discuss the solution of equation (\ref{2.2}) 
below, but now we continue the construction of the model. We place a 
quantum spin $s_l$ at each vertex of the graph. We construct the 
Hamiltonian describing the interaction of nearest neighbors $\langle 
kl\rangle$:
\begin{equation}
H=\sum_{\langle kl\rangle}H(k,l). \label{2.3}
\end{equation}
Here the summation goes over the edges. The Hamiltonian density is
\begin{equation}
H(k,l)=\sum_{J=s_k+s_l+1-M_{kl}}^{s_k+s_l}A_J(k,l)\Pi_J(k,l). \label{2.4}
\end{equation}
Here $\Pi_J$ is the projection (\ref{1.8}) while the coefficients 
$A_J(k,l)$ are real and positive--parameters of the model (depending on 
$J$ and on the edge). The model is thus determined. In analogy to \S 1 we 
prove the existence and uniqueness of an eigenfunction describing the 
ground state. This VBS state is such that
$$H|\psi\rangle =0,
$$
where
\begin{equation}
|\psi\rangle =\prod_{\langle 
kl\rangle}(a^+_kb^+_l-a^+_lb^+_k)^{M_{kl}}|0\rangle . \label{2.5}
\end{equation}

We anticipate that Hamiltonian (\ref{2.4}) has a gap in the spectrum; the 
correlators (see \S 11) decay exponentially for quasi periodic coverings, 
$M_{kl}\ge 1$.

Several subsequent sections are devoted to the solution of equations 
(\ref{2.1}) and (\ref{2.2}). The explicit form of the wave function 
(\ref{2.5}) shows that nullification of any of the numbers $M_{kl}$ is 
equivalent to the absence of an edge. It is therefore a question of 
solving system (\ref{2.1}) in positive integers $M_{kl}\ge 1$. Of course, 
the integers $M_{kl}$ can be prescribed arbitrarily, and the values of 
the spins can be computed. It is, however, interesting to solve the 
inverse problem and clarify what restrictions the system (\ref{2.2}) 
imposes on the permitted values of the spins. For example, it is clear 
that the spin can be equal to 1/2 only when a vertex has one nearest 
neighbor. We shall consider only connected graphs. For disconnected 
graphs the problem reduces to several independent problems.

\vskip 0.5cm
{\bf \S 3. Linear graphs.}
\vskip 0.5cm

We shall first solve system (\ref{2.2}) for linear graphs. We consider a 
non closed, one--dimensional chain with $N$ nodes. Let $l$ be the index of 
a node, $l=1,2,\ldots ,N$. We complete the relations (\ref{2.1}) at the 
ends as follows:
\begin{equation}
M_{01}=0 \ {\rm and} \ M_{N,N+1}=0. \label{3.1}
\end{equation}
A solution of equation (\ref{2.1}) has the form
\begin{equation}
M_{l,l+1}=\sum_{k=1}^l(-1)^{k-l}\cdot 2s_k. \label{3.2}
\end{equation}
For $l=N$ it follows from this that
\begin{equation}
\sum_{k=1}^N(-1)^k\cdot 2s_k=0. \label{3.3}
\end{equation}
>From the positivity of $M$ it follows that
\begin{equation}
M_{l,l+1}=\sum_{k=1}^l(-1)^{k-l}\cdot 2s_k\ge 1. \label{3.4}
\end{equation}

We now consider a periodic chain with an even number of nodes $N$. The 
equation
\begin{equation}
2s_l=M_{l,l+1}+M_{l-1,l} \label{3.5}
\end{equation}
has a non unique solution. A solution of the homogeneous equation
\begin{equation}
0=M_{l,l+1}+M_{l-1,l} \label{3.6}
\end{equation}
has the form
\begin{equation}
M_{l,l+1}=(-1)^l\alpha . \label{3.7}
\end{equation}
The nodes with indices 1 and $N+1$ are identified. This can be used to 
break the chain, i.e., to set, for example,
\begin{equation}
M_{1,N}=0. \label{3.8}
\end{equation}
This reduces the problem to an unclosed chain, which has already been 
solved. Thus, a solution exists only in the case
\begin{equation}
\sum_{l=1}^N(-1)^l\cdot 2s_l=0. \label{3.9}
\end{equation}
It has the form
\begin{equation}
M_{l,l+1}+(-1)^{l+1}M_{1,N}=\sum_{k=1}^l(-1)^{k-l}\cdot 2s_k. \label{3.10}
\end{equation}
Here $M_{1,N}$ is an arbitrary positive integer. For odd $l$ the left 
side of (\ref{3.10}) is positive, which imposes restrictions on the right 
side of (\ref{3.10}). Thus, recalling positivity, it is possible to 
change $\{ M_{l,l+1}\}$ for fixed values of the spin $s$. This means that 
for the same choice of spins it is possible to produce several distinct 
Hamiltonians (\ref{2.4}) whose definition contains the collection $\{ 
M_{l,l+1}\}$. Each of these Hamiltonians will have a unique eigenfunction 
of the ground state.

We now consider a cycle with an odd number of nodes. Equation (\ref{3.5}) 
has the unique solution
\begin{equation}
M_{l,l+1}=\sum_{k=1}^l(-1)^{l-k}\cdot 2s_k-\sum_{k=1}^N(-1)^{l-k}\cdot 
s_k\ge 1. \label{3.11}
\end{equation}
>From this it is clear that the sum 
$\ds\sum_{k=1}^N(-1)^k\cdot s_k=(-1)^NM_{1,N}$ must be an integer. Since we 
have the equality
\begin{equation}
\sum_{k=1}^Ns_k+\sum_{k=1}^N(-1)^k\cdot s_k=2\sum_{k\equiv 0~(\mod 2)}s_k, 
\label{3.12}
\end{equation}
it follows that the sum $\ds\sum_{k=1}^Ns_k$ is an integer. Hence,
\begin{equation}
\sum_{k=1}^N2s_k={\rm an~even~number}. \label{3.13}
\end{equation}

\vskip 0.5cm
{\bf \S 4. Quasicrystals.}
\vskip 0.5cm

There are now a large number of works devoted to quasicrystals (see, for 
example, [11]--[14]. Analysis of dynamical system in quasicrystals is 
also of broad interest. For example, it has been possible to solve the 
Ising model and the eight--vertex Baxter model in two--dimensional 
quasicrystal [14].

In connection with the problem of finding the densest packing specific 
 associated with quasicrystals are of major interest [15].

Here we shall construct a generalized model of a quantum antiferromagnet 
in a quasicrystal of two or higher number of dimensions. The simplest 
example of a quasi periodic covering is the Penrose tiling of the plane by 
translations of ten rhombuses. Corresponding figures can be found in 
[13], [14]. In the situation of general position a quasicrystal (its 
vertices and edges) is bipartite graph [16].

An important characteristic of a quasicrystal is the number of nearest 
neighbors of vertices (the coordination number $z_l$). It can run through 
only a finite number of values. For example, for the Penrose tiling of 
the plane by translations of 10 rhombuses we have [15] $3\le z_l\le 7$. 
For filling out three--dimensional space by translations of 20 
parallelepipeds of special form (rhomboids) we have [15] $4\le z_l\le 
12$. This filling has the symmetry of a right icosahedron and realizes 
the crystal structure of the rapidly cooled alloy ${\rm Al}_6{\rm Mn}$ 
[17], [18]. A generalized quantum antiferromagnet can be constructed in a 
quasicrystal in the same way as on an arbitrary graph. The general 
outline was described above. We shall present only the simplest example.

At the vertices of the quasicrystal we place spins which, generally 
speaking, are different in magnitude. The spin situated at a given vertex 
is equal to half the correlation number:
\begin{equation}
s_l=\frac{1}{2}z_l. \label{4.1}
\end{equation}
For Penrose rhombuses $3/2\le s_l\le 7/2$, while for the icosahedral 
filling of space by rhomboids $2\le s_l\le 6$. The interaction 
Hamiltonian for these spins can be taken, for example, in the form
\begin{equation}
H=\sum_{\langle kl\rangle}\Pi_{s_k+s_l}(k,l). \label{4.2}
\end{equation}
The Hamiltonian density is the projection onto the highest possible value 
of the spin arising in adding spins at two neighboring nodes $k$ and $l$. 
We note that the model constructed here of a quantum antiferromagnet in a 
quasicrystal has a unique ground state (the valence bond state), which 
differs from the N\'eel state.

A quasicrystal is a quasi periodic filling of all space. An important 
characteristic of it is the following. Any finite part of quasicrystal 
has an infinite set of copies, and these copies repeat throughout space 
with a particular probability. It is just this that makes it possible to 
justify the presence of a thermodynamic limit in quasicrystal. In analogy 
to [5] it is possible to show that there is a gap in the spectrum of the 
Hamiltonian (\ref{4.2}), and the correlators of the spins decay 
exponentially. In analogy to [4] it can be shown that the model of 
quantum antiferromagnet in quasicrystal is equivalent to a model of 
classical statistical physics in the same quasicrystal with Hamiltonian 
density $-\log ((1-{\bf n}_k{\bf n}_l)/2)$. Here ${\bf n}$ is a unit vector on 
the sphere.

\vskip 0.5cm
{\bf \S 5. Tree graphs.}
\vskip 0.5cm

For a disconnected graph the system of equations (\ref{2.1}) decomposes 
into several independent systems. Therefore we henceforth consider only 
connected graphs. The procedure of "cutting-off branches" is useful in 
the investigation of the system (\ref{2.1}) for an arbitrary graph; we 
proceed to a description of it. Here an important role is played by 
bipartite graphs. by definition, the set of vertices $\{ l\}$ of 
bipartite graph can be broken into two non intersecting subsets $\{ l\}=\{ 
A\}\cup\{ B\}$ so that an edge of the initial graph joins only vertices 
of different subsets. To each vertex $l$ of the graph $\Gamma$ it is 
possible to ascribe a parity
\begin{equation}
\epsilon_l=\left\{\begin{array}{ll} 1,& {\rm if} \ l\in A,\\
-1,& {\rm if} \ l\in B. \end{array}\right. \label{5.1}
\end{equation}
We now consider a graph which decomposes into two disconnected subgraphs 
$\Gamma_1$ and $\Gamma_2$ when one edge is removed. here $\Gamma_1$ is a 
bipartite graph. we denote vertices belonging to the edge $b$ by $a$ and 
$c$ ($a\in\Gamma_1$, $c\in\Gamma_2$). It is easy to compute the integer 
$M$ corresponding to the edge $b$:
\begin{equation}
M_b=\sum_{l\in\Gamma_1}\epsilon_l\cdot 2s_l. \label{5.2}
\end{equation}
The contributions of the edges of the bipartite graph cancel on the right 
side of (\ref{5.2}). The signs are chosen so that $\epsilon_a=1$. We now 
redefine the magnitude of the spin at the nodes $a$ and $c$:
\begin{equation}
2s_a\mapsto 2s_a-M_b; \ \ 2s_c\mapsto 2s_c-M_b. \label{5.3}
\end{equation}
Thus, solution of equation (\ref{2.2}) for the entire graph has reduced 
to the solution of two independent equations of the type (\ref{2.2}) for 
each subgraphs $\Gamma_1$  and $\Gamma_2$ which are not connected with one 
another. This procedure of reducing the system (\ref{2.1}) to two simpler 
independent systems is naturally called the procedure of "cutting--off 
branches". We apply this procedure to the investigation of system 
(\ref{2.1}) foe the tree graphs. Thus, we consider the graphs without 
cycles (tree graphs). We denote them by the letter ${\cal D}$. It is well 
known that any tree is a bipartite graph [10]. We shall study equation 
(\ref{2.2}) in this case. We consider any edge $\langle kl\rangle$. 
Removing it leads to a decomposition of the tree into two disconnected 
trees ${\cal D}_1\cup{\cal D}_2$ where $k\in{\cal D}_1$, $l\in{\cal D}_2$.

We choose an alternating sum of spins along the tree ${\cal D}_1$, so 
that $s_k$ enters with sign +1. It is clear that
\begin{equation}
M_{kl}=\sum_{p\in{\cal D}_1}\epsilon_p\cdot 2s_p. \label{5.4}
\end{equation}
This is a solution of equation (\ref{2.2}). it is also possible to 
express $M_{kl}$ in terms of an analogous sum over the second tree. 
Comparison  of the two expressions leads to the relation
\begin{equation}
2\sum_{l\in{\cal D}}\epsilon_p\cdot 2s_p. \label{5.5}
\end{equation}
Moreover, the following inequality must be satisfied (the condition of 
positivity):
\begin{equation}
M_{kl}=2\sum_{p\in{\cal D}_1}\epsilon_ps_p\ge 1. \label{5.6}
\end{equation}

The Cayley tree is a particular example of a tree.

We note that in [5] the case where all spins $s=3/2$ is considered for a 
finite Cayley tree (with $z=3$). It is easy to see that then the system 
of equations (\ref{2.2}) has positive integer solutions. The spin must be 
equal to 3/2 only for interior nodes, while on the boundary $s=1/2$. This 
guarantees the uniqueness theorem.

We note that it is just the requirement of positivity which does not 
permit taking all spins equal to one another for a tree of general 
position.

\vskip 0.5cm
{\bf \S 6. Bipartite connected graphs. A criterion of positivity.}
\vskip 0.5cm

We remark that if there is a cycle of even length in an arbitrary graph, 
then a solution of equation (\ref{2.2}) is not unique. Indeed, along a 
cycle it is always possible to add to the quantities $M$ a quantity 
$(-1)^l\alpha$ analogous to (\ref{3.7}). It does not change $s_l$. The 
quantity $\alpha$ can be chosen so that one of the quantities $M$ 
vanished. This corresponds to removing an edge and breaking the cycle. 
Thus, all even cycles can be broken without changing the spins but by 
changing $M$. For a bipartite graph any cycle is even [10]; therefore, 
the problem on a bipartite graph reduces to the problem on a tree graph. 
Hence, equation (\ref{2.2}) is solvable in integers if and only if
\begin{equation}
\sum\epsilon_ls_l=0. \label{6.1}
\end{equation}
The positivity condition $M_{kl}\ge 1$ imposes more complicated 
restrictions on the spin. An obvious consequence of (\ref{2.1}) is the 
condition
$$2s_l\ge z_l.
$$
Here $z_l$ is the coordination number of the $l$th node.

For bipartite graphs we shall derive a more refined necessary condition. 
For this we make several definitions. We denote our graph by $\Gamma$. 
Suppose that by cutting $n$ edges it can be broken into two independent 
subgraphs $\Gamma_1$ and $\Gamma_2$. Suppose that the following 
conditions is thereby satisfied. All vertices belonging simultaneously to 
$\Gamma_1$ and the cut edges belong only to one sub lattice, for example, 
$A$. We chose an alternating sum of spins along the subgraph $\Gamma_1$:
\begin{equation}
\sum_{l\in\Gamma_1}\epsilon_l\cdot 
2s_l=\sum_{l\in\Gamma_1}\sum_{k\in\Gamma_2}M_{lk}. \label{6.2}
\end{equation}
Here $\epsilon_l=1$ for the sub lattice $A$ and $\epsilon_l=-1$ for the 
sub lattice $B$. The sum of $M_{kl}$ along the cut edges stands on the 
right side of (\ref{6.2}).

>From (\ref{6.2}) it is evident that
\begin{equation}
\sum_{l\in\Gamma_1}\epsilon_l\cdot 2s_l\ge n. \label{6.3}
\end{equation}
Here $n$ is the number of cut edges. This condition is necessary. We have 
been unable to prove sufficiency of this condition.

\vskip 0.5cm
{\bf \S 7. Non bipartite, connected graphs.}
\vskip 0.5cm

It was shown in \S 6 that the presence of an even cycle in a graph leads 
to non uniqueness of the system (\ref{2.2}). It turns out that two odd 
cycles joined by a chain lead to analogous degeneracy.
\vskip 0.5cm

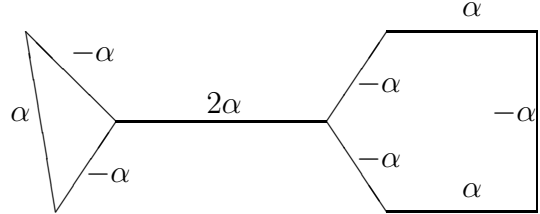
\setlength{\unitlength}{0.4cm}
\begin{picture}(16,9)(-7,-1)
\put(1,0){\line(-1,6){1}}
\put(1,0){\line(2,3){2}}
\put(3,3){\line(-1,1){3}}
\put(3,3){\line(1,0){7}}
\put(10,3){\line(2,-3){2}}
\put(10,3){\line(2,3){2}}
\put(12,0){\line(1,0){5}}
\put(12,6){\line(1,0){5}}
\put(17,0){\line(0,1){6}}
\put(6,3.3){\hbox{$2\alpha$}}
\put(-0.5,3){\hbox{$\alpha$}}
\put(2,1){\hbox{$-\alpha$}}
\put(1.5,5){\hbox{$-\alpha$}}
\put(11,1.5){\hbox{$-\alpha$}}
\put(11,4){\hbox{$-\alpha$}}
\put(14.5,0.5){\hbox{$\alpha$}}
\put(14.5,6.5){\hbox{$\alpha$}}
\put(15.5,3){\hbox{$-\alpha$}}
\end{picture}
\vskip 0.5cm

The scheme shown admissible changes of the number $M$ (corresponding to 
edges) which do not change the spins. (Indeed, the sum of the changes at 
each node is equal to 0.) It suffices that the graph shown in the scheme 
be subgraph of $\Gamma$; this already implies degeneracy of the system 
(\ref{2.2}) on $\Gamma$. It is easy to see that the graph in the scheme 
was obtained from an even cycle (of length 10) by gluing together two 
sides. This is a degenerate even cycle. Thus, we consider the system 
(\ref{2.2}) on a non bipartite graph. We begin to simplify it. We first 
break all even cycles. We then remove all degenerate even cycles so that 
the connectivity of the graph is preserved. For this we remove only 
non degenerate edges in the operation of breaking cycles in the degenerate 
case. Under this method of breaking degenerate cycles the connectivity of 
the graph is preserved [10]. In summary we break all degenerate even 
cycles. After applying the procedure of "cutting--off branches" (see \S
6) we arrive at a connected graph consisting of a single odd cycle. For 
one odd cycle the problem has already been solved (see (\ref{3.11}). The 
condition for solvability in integers is the parity of the sum $\sum 
2s_k$ (see (\ref{3.13})). The condition of positivity of $M$ requires 
further study.

\vskip 0.5cm
{\bf \S 8. The modified classical Heisenberg model.}
\vskip 0.5cm

In [6] it was shown that the quantum Heisenberg model considered in \S \S 
1,2 is equivalent to a modified classical Heisenberg model. This model 
can be described as follows. The three--components unit vector ${\bf 
n}(l)$ (classical) is situated at the $l$th node of the lattice. The 
statistical sum of the classical model is
\begin{equation}
\Phi =\int_{S^2}\prod_{l=-L}^Ld\Omega_l\prod_{l=-L}^{L-1}
\left(\frac{1-{\bf n}(l){\bf n}(l+1)}{2}\right)^{M(l)} \label{8.1}
\end{equation}
and is equal to the square of the norm of the VBS wave function
\begin{equation}
|\psi\rangle 
=\prod_{l=-L}^{L-1}(a^+_lb^+_{l+1}-a^+_{l+1}b^+_l)^{M(l)}|0\rangle . 
\label{8.2}
\end{equation}

The multi point correlation functions are computed by the formula
\begin{eqnarray}
\langle\psi |~\prod_{j=1}^N\wh S^{a_j}(r_j)~|\psi\rangle &=&
\prod_{j=1}^N(s(l_j)+1)\int_{S^2}\prod_{l=-L}^Ld\Omega_l \label{8.3}\\
&\times&\prod_{l=-L}^L\left(\frac{1-{\bf n}(l)\cdot{\bf 
n}(l+1)}{2}\right)^{M(l)}\prod_{j=1}^N{\bf n}^{a_j(r_j)}. \nonumber
\end{eqnarray}
Here $r_N>r_{N-1}>\cdots >r_2>r_1$, $s(l)$ in the magnitude of the spin 
at the $l$th node, and $n^a(l)$ is the component with index $a$ 
($a=1,2,3$) of the vector ${\bf n}(l)$.

We introduce the integral operator $\wh K_M$. It acts in the space of 
functions $f({\bf n})$ on the unit sphere $S^2$ according to the formula
\begin{equation}
(\wh K_Mf)({\bf n}_2)=\int_{S^2}d\Omega\left(\frac{1-{\bf n}_1\cdot{\bf n}_2}
{2}\right)^Mf({\bf n}_1). \label{8.4}
\end{equation}
In the next section we diagonalize the operator $\wh K_M$ and show that 
the operators $\wh K_M$ from a commuting family. An individual operator 
$\wh K_M$ is linear combination of $M+1$ projections. Using the 
properties of the operators $\wh K_M$, we compute in explicit form the 
norm of the function (\ref{8.2}) and the correlation functions.

We note that the VBS wave function of the inhomogeneous Heisenberg model 
for an arbitrary graph was found in (\ref{2.5}), \S 2. The statistical 
sum for the modified classical Heisenberg model is equal to the norm of 
the VBS wave function (\ref{2.5}) and can be computed by the formula
\begin{equation}
\Phi =\langle\psi |\psi\rangle =\int_{S^2}\prod_ld\Omega_l
\prod_{\langle kl\rangle}\left(\frac{1-{\bf n}_k\cdot{\bf 
n}_l}{2}\right)^{M_{kl}}. \label{8.5}
\end{equation}
Here $M_{kl}\ge 0$ is an arbitrary choice of alternating numbers. For the 
complete graph all the vertices are joined by edges. If for some edge 
$\langle kl\rangle$ we have $M_{kl}=0$, then this is equivalent to the 
absence of the edge $\langle kl\rangle$ in the graph. Hence, we can 
consider an arbitrary graph as special case of a complete graph.

\vskip 0.5cm
{\bf \S 9. Commuting transfer matrices.}
\vskip 0.5cm

We consider the integral operator $\wh K_M$ with kernel
\begin{equation}
K_M({\bf n}_1,{\bf n}_2)=\left(\frac{1-{\bf n}_1\cdot{\bf 
n}_2}{2}\right). \label{9.1}
\end{equation}
We shall first find its eigenfunction. We denote them by 
$\Lambda_M^A({\bf n})$ -- these are symmetric, traceless tensors of rank 
$N$. For example,
\begin{eqnarray}
&&\Lambda_0=1, \ \ \Lambda_1^a=n^a, \ \ \Lambda_2^{a_1a_2}=n^{a_1}n^{a_2}-
\frac{1}{3}\delta^{a_1}_{a_2},\nonumber \\
&& \label{9.2} \\
&&\Lambda_3^{a_1a_2a_3}=n^{a_1}n^{a_2}n^{a_3}-\frac{1}{5}(n^{a_1}
\delta^{a_2}_{a_3}+n^{a_2}\delta^{a_1}_{a_3}+n^{a_3}\delta^{a_1}_{a_2}).
\nonumber 
\end{eqnarray}
Before writing out the general formula for $\Lambda^A({\bf n})$, it is 
useful to introduce some notation. For a collection of indices 
$A_N=\{a_j\}$ we set
\begin{equation}
n^{A_N}=\prod_{j=1}^Nn^{a_j}. \label{9.3}
\end{equation}
For even $N$, $N=2k$, an important role is played below by a partition of 
the set $A_N$ into pairs $A_{2k}=\cup_{\alpha}\{a_{\alpha}b_{\alpha}\}$. 
We define the delta function of the set $A_N$ by the formula
\begin{equation}
\delta (A_{2k})=\frac{1}{k!}\sum_{A_{2k}=\cup_{\alpha}\{ a_{\alpha}
b_{\alpha}\}}\prod_{\alpha}\delta^{a_{\alpha}}_{b_{\alpha}}. \label{9.4}
\end{equation}
We note that the number of terms in (\ref{9.4}) is equal to
$$(2k-1)!!=(2k)!/(2^k\cdot k!).
$$
\begin{th} \label{t9.1} The eigenfunction $\Lambda^{A_N}$ of the 
integral operator $\wh K_M$ are given by the formula
\begin{equation}
\Lambda^{A_N}({\bf n})=\sum_{k=0}^{[N/2]}\gamma_k(N)\sum_{A_{2k}\subset A_N}
n^{A_N/A_{2k}}\delta(A_{2k}). \label{9.5}
\end{equation}
Here $(A_N/A_{2k})\cup A_{2k}=A_N$. The coefficient $\gamma_k(N)$ is
\begin{equation}
\gamma_k(N)=\prod^k_{j=1}\left(\frac{-1}{2N-2j+1}\right); \ \ \gamma =1. 
\label{9.6}
\end{equation}
The tensor $\Lambda^{A_N}$ is a polynomial in $n^a$ whose leading 
component is equal to $n^{A_N}$.
\end{th}
\begin{th} \label{t2.1} The polynomial $\Lambda^{A_N}({\bf n})$ is an 
eigenfunction of the operator $\wh K_M$,
\begin{equation}
\int_{S^2}d\Omega_l\left(\frac{1-{\bf n}_1\cdot{\bf n}_2}{2}\right)^M
\Lambda^{A_N}({\bf n}_1)=z(N,M)\Lambda^{A_N}({\bf n}_2), \label{9.7}
\end{equation}
with eigenvalue
\begin{equation}
z(N,M)=\frac{1}{M+1}\prod_{j=0}^{N-1}\frac{j-M}{j+2+M}, \ \ 
z(0,M)=\frac{1}{M+1}. \label{9.8}
\end{equation}
\end{th}

It is interesting to note that
$$z(N,M)=0, \ \ {\rm if} \ N\ge M+1.
$$

Hence, the integral operator $\wh K_M$ is a linear combination of $M+1$ 
projections. The details of the computation of $z(N,M)$ are presented in 
Appendix~B. We note that the $\Lambda^{A_N}({\bf n})$ are eigenfunction 
of the Laplace operator on the sphere $S^2$ (for details see \S 12).

The  of the operator $\wh K_M$ do not depend on $M$ (see 
(\ref{9.5})). This implies that the operators $\wh K_M$ commute for 
different values of $M$:
$$[{\wh K}_{M_1},{\wh K}_{M_2}]=0.
$$

The norm of the VBS wave function or the statistical sum for the modified 
Heisenberg model (\ref{8.1}) and the correlation functions (see \S 10) 
can be expressed in terms of the transfer matrices
\begin{equation}
T(r_2,r_1)=\prod_{l=r_1}^{r_2-1}{\wh K}_{M(l)}. \label{9.9}
\end{equation}
All factors in (\ref{9.9}) commute with one another. The set of 
eigenfunctions of the transfer matrix $T$ is given by (\ref{9.5}) and 
does not depend on the collection of numbers $\{ M(l)\}$. Hence, the 
transfer matrices (\ref{9.9}) form a commuting family for different 
collections of the numbers $\{ m(l)\}$.

The number of independent components of the tensor $\Lambda_N^{A_N}$ 
(which is symmetric and traceless) is equal to $2N+1$. The tensor 
$\Lambda_N^{A_N}$ generates an irreducible representation of the algebra 
$o(3)$ of spin $N$. The product of irreducible representations decomposes 
into a direct sum of irreducible representations. We shall need the 
explicit form of this decomposition in the case where one of the spins is 
equal to 1:
\begin{eqnarray}
\Lambda_N^{A_N}\cdot n^{a_{N+1}}&=&\Lambda_{N+1}^{A_N\cup a_{N+1}}+
\frac{1}{2N+1}\sum_{j=1}^N\Lambda_{N-1}^{a_1\ldots\wh a_j\ldots a_N}
\delta^{a_j}_{a_{N+1}}\\
&-&\frac{2}{(2N+1)(2N-1)}\sum_{1\le i<j\le N}\Lambda_{N-1}^{a_1\ldots 
\wh a_i\ldots\wh a_j\ldots a_{n+1}}\delta_{a_j}^{a_i}.\nonumber \label{9.10}
\end{eqnarray}
We recall that $A_N=\{ a_l,\ldots ,a_N\}$. We note that in adding spin 
$N$ and spin 1 only the spins $N+1$ and $N-1$ occur. We further define 
the projections $P$ onto spin $N+1$ and spin $N-1$ by the formulas
\begin{equation}
P_{N+1}(\Lambda_N^{A_N}\cdot n^{a_{N+1}})=\Lambda_{N+1}^{A_N\cup 
a_{N+1}}, \label{9.11}
\end{equation}
\begin{eqnarray}P_{N-1}(\Lambda_n^{A_N}\cdot n^{a_{N+1}})&=&\frac{1}{2N+1}
\sum_{j=1}^N\Lambda_{N-1}^{a_1\ldots\wh a_j\ldots 
a_N}\delta^{a_j}_{a_{N+1}} \label{9.12}\\
&-&\frac{2}{(2N+1)(2N-1)}\sum_{1\le i<j\le N}\Lambda_{n-1}^{a_1\ldots
\wh a_i\ldots\wh a_j\ldots a_{N+1}}\delta_{a_j}^{a_i}. \nonumber
\end{eqnarray}
We remark that $P_{N-1}+P_{N+1}=1$ which follows from (\ref{9.10}). The 
value of the projections with other indices on the polynomial 
$\Lambda_N^{A_N}{\bf n}^{a_{N+1}}$ we define to be zero:
$$P_J(\Lambda_N^{A_N}n^{a_{N+1}})=0, \ \ {\rm if} \ j\ne N\pm 1.
$$

It is natural to consider the number $N$ to be the spin of the tensor 
$\Lambda^{A_N}$. In computing multi point correlators it turns out to be 
useful to consider the following combination of projections:
\begin{equation}
P_0(n^{a_N}P_{k_{N-1}}(n^{a_{N-1}}P_{k_{N-2}}(n^{a_{N-2}}\cdots
P_{k_4}(n^{a_4}P_{k_3}(n^{a_3}P_{k_2}n^{a_2a_1})))\cdots ). \label{9.13}
\end{equation}
The expression (\ref{9.13}) is well defined and is a linear combination 
of products of the delta functions $\delta_{a-k}^{a_j}$. It is different 
from 0 only if $k_2=0,2$ and for $j=2,\ldots ,N-2$ the equalities 
$|k_{j+1}-k_j|=1$ are satisfied. We present some examples:
$$P_0(n^{a_1a_2})=\frac{1}{3}\delta^{a_1}_{a_2}, \ 
P_2(n^{a_1a_2})=\Lambda_2^{a_1a_2}, \ P_0(n^{a_3}P_k(n^{a_1a_2}))=0,
$$
\begin{equation}
P_1(\Lambda_2^{a_1a_2}n^{a_3})=\frac{1}{5}(n^{a_1}\delta^{a_2}_{a_3}+
n^{a_2}\delta^{a_1}_{a_3})-\frac{2}{15}n^{a_3}\delta_{a_2}^{a_1}, 
\label{9.14}
\end{equation}
$$P_0(n^{a_4}P_1(n^{a_3}P_2(n^{a_1a_2})))=\frac{1}{15}(\delta^{a_1}_{a_4}
\delta_{a_3}^{a_2}+\delta_{a_3}^{a_1}\delta_{a_4}^{a_2})-
\frac{2}{45}\delta_{a_2}^{a_1}\delta_{a_4}^{a_3}.
$$
A combination of the projections (\ref{9.13}) is a collection of 
nonnegative integers $\{ k_0,k_1,k_2,\ldots ,k_{N-1},k_N\}$ such that 
$k_0=0$, $|k_{i+1}-k_i|=1$, $i=0,1,\ldots , N-1$, $k_N=0$. It is clear 
that it is also possible to consider such collections with another 
boundary condition $k_N=l$. It is obvious that $l\le N$, $l\equiv N~(\mod 
2)$. In \S 11 we show that for fixed $k_N:=l$ such sequences can be 
parameterized by standard Young tableaux of the form $((N+l)/2,(N-l)/2)$. 
Hence, the number of different combinations of projections of the form 
(\ref{3.13}) is equal to 0  for $N$ odd and equal to the Catalan number 
$C_k=(2k)!/k!(k+1)!$ if $N=2k$.

We finish this section with the remark that in principle, using the rules 
of passing from functions of spin variables to functions of spherical 
coordinates on the sphere $S^2$ described in [19], it is possible to find 
an expression for $\Lambda^{A_N}$ in terms of the original quantum spins 
$\wh S^a$ (see \S 1). We illustrate the nature of the answer only with 
one example. We consider 
$\Lambda_2^{a_1a_2}=n^{a_1a_2}-\frac{1}{3}\delta_{a_2}^{a_1}$. In terms 
of the quantum spins $\Lambda_2$ can be written as follows:
$$\Lambda_2^{a_1a_2}=\frac{\wh S^{a_1}\wh S^{a_2}+\wh S^{a_2}\wh S^{a_1}-
\frac{2}{3}s(s+1)\delta^{a_1}_{a_2}}{(s+1)(2s+3)}.
$$
\vskip 0.5cm
{\bf \S 10. The norm and the two--point correlator.}
\vskip 0.5cm

We first consider an open chain and compute the norm of the VBS wave 
function (\ref{8.2}). We shall proceed from formula (\ref{8.1}). We 
remark that it is possible to rewrite the expression for the square of 
the norm of the wave function in terms of the transfer matrix (\ref{9.9}):
\begin{eqnarray}
\langle\psi ~|~\psi\rangle &=&\int\prod^L_{l=-L}d\Omega_l\prod_{l=-L}^{L-1}
\left(\frac{1-{\bf n}(l)\cdot{\bf n}(l+1)}{2}\right)^{M(l)}\nonumber \\
&& \label{10.1} \\
&=&\psi_0=\Lambda_0\prod_{l=-L}^{L-1}\wh K_{M(l)}\Lambda_0. \nonumber
\end{eqnarray}
Using (\ref{9.7}) and (\ref{9.8}), we find
\begin{equation}
\langle\psi ~|~\psi\rangle_0=\prod_{l=-L}^{L-1}\frac{1}{M(l)+1}. 
\label{10.2}
\end{equation}
The norm of the wave function for a periodic chain can be computed in a 
similar way. Indeed,
\begin{equation}
\langle\psi ~|~\psi\rangle_{\rm reg}=\tr\prod_{l=-L}^{L-1}\wh K_{M(l)}=
\tr T(L,-L). \label{10.3}
\end{equation}
All the operators $\wh K$ commute, and their spectrum is known. The 
transfer matrix $T(L,-L)$ has $m+1$ nonzero eigenvalues. Here 
$m=\min\{ M(l)\}$. According to (\ref{9.8}), these eigenvalues are equal to
\begin{equation}
Z_N=\prod_{l=-L}^{L-1}Z(N,M(l)), \ \ N=0,1,\ldots ,m. \label{10.4}
\end{equation}
Hence, the square of the norm of the wave function for a periodic chain is
\begin{equation}
\langle\psi ~|~\psi\rangle_{\rm reg}=\sum_{N=0}^m(2N+1)\cdot Z_N. 
\label{10.5}
\end{equation}
The factor $2N+1$ in (\ref{10.5}) describes the multiplicity of the 
degeneracy of the eigenvalue $Z_N$.

An especially simple formula for $\langle\psi ~|~\psi\rangle_{\rm reg}$ 
is obtained for $m=1=M(L)$:
$$\langle\psi ~|~\psi\rangle_{\rm reg}=\prod_{l=-L}^{L-1}\frac{1}
{M(l)+1}-\frac{1}{2}\prod_{l=-L}^{L-1}\frac{-M(L)}{(M(l)+1)(M(l)+2)}.
$$
We now consider the thermodynamic limit for $\langle\psi 
~|~\psi\rangle_{\rm reg}$ as $L\to\infty$. It is easy to see that
\begin{equation}
\langle\psi ~|~\psi\rangle_{\rm reg}=\sum_{N=0}^m(2N+1)Z_N\to Z_0=
\prod_{l=-L}^{L-1}\frac{1}{M(l)+1}=\langle\psi ~|~\psi\rangle_0. 
\label{10.6}
\end{equation}
Hence, the thermodynamic limit does not depend on the boundary 
conditions. Computation of the limit in (\ref{10.6}) is based on the 
inequality
\begin{equation}
Z(N_2,M)<Z(N_1,M) \ \ {\rm if} \ N_1<N_2. \label{10.7}
\end{equation}

We now proceed to the computation of the two--point correlator for an 
open chain. We use the representation (\ref{8.3}):
\begin{equation}
\psi_0^{-1}\int\prod_ld\Omega_l\prod_{l=-L}^{L-1}\left(\frac{1-{\bf n}(l)
\cdot{\bf n}(l+1)}{2}\right)^{M(l)}{\bf n}^{a_1}(r_1)\cdot{\bf n}^{a_2}(r_2). 
\label{10.8}
\end{equation}
Here $r_2>r_1$. Formula (\ref{10.8}) for the correlator can be written in 
terms of the transfer matrix (\ref{9.9}) in the following manner:
\begin{equation}
\psi_0^{-1}\cdot\Lambda_0\cdot T(L,r_2)n^{a_2}(r_2)T(r_2,r_1)
n^{a_1}(r_1)T(r_1,-L)\Lambda_0. \label{10.9}
\end{equation}
In (\ref{10.9}) the transfer matrices are defined in analogy to 
(\ref{9.9}). For example,
\begin{equation}
T(L,r_2)=\prod_{l=r_2}^{L-1}\wh K_{M(l)}, \ \ r_2<L. \label{10.10}
\end{equation}
The function $\Lambda_0$ is an eigenfunction for the transfer matrix 
$T(r_1,-L)$ with eigenvalue
\begin{equation}
\prod_{l=-L}^{r_1-1}\frac{1}{M(l)+1}. \label{10.11}
\end{equation}
It is easy to see that (\ref{10.11}) is contained as a factor in 
$\langle\psi ~|~\psi\rangle_0$ and hence cancels in (\ref{10.9}). The 
function $n^{a_1}(r_1)=\Lambda_1^{a_1}$ is an eigenfunction for 
(\ref{10.10}) with eigenvalue
\begin{equation}
\prod_{l=r_1}^{r_2-1}\frac{-M(l)}{(M(l)+1)(M(l)+2)}. \label{10.12}
\end{equation}
Further, it is clear that $T(L,r_2)$ does not contribute to the 
correlator. Finally, for (\ref{10.8}) we find
\begin{equation}
\langle n^{a_2}(r_2)n^{a_1}(r_1)\rangle =\frac{1}{3}\delta^{a_1}_{a_2}
\prod_{l=r_1}^{r_2-1}\left(\frac{-M(l)}{M(l)+2}\right). \label{10.13}
\end{equation}
Formula (\ref{10.13}) gives an expression for two--point correlator for 
finite open chain. Using (\ref{8.3}), we find the correlator of two spins
\begin{equation}
\langle S^{a_1}(r_1)S^{a_2}(r_2)\rangle =\frac{(s(r_1)+1)(s(r_2)+1)}{3}
~\delta^{a_1}_{a_2}\prod_{l=r_1}^{r_2-1}\left(\frac{-M(l)}{M(l)+2}\right) 
.\label{10.14}
\end{equation}
Formula (\ref{10.14}) gives an expression for the two--point correlator 
for the quantum model (\ref{2.4}). In the thermodynamic limit formula 
(\ref{10.4}) is preserved (the thermodynamic limit does not depend on the 
boundary conditions). It also gives an answer for the one--dimensional 
quasicrystal (\ref{0.10}) in which $M(l)$ assumes only the two values 
$M_s$ and $M_l$ in a quasi periodic manner. The asymptotic (\ref{0.12}) 
follow directly from (\ref{10.14}). If all the numbers $M(l)$ are equal, 
then $M=s$ and (\ref{10.14}) reproduces the result (\ref{0.8}).

We now rewrite formula (\ref{10.15}) for the square of the norm of the 
VBS wave function in terms of generalized hyper geometric functions 
$~_pF_q$. We recall their definitions (see [20]):
\begin{equation}
~_pF_q\left(\begin{array}{c}\alpha_1,\ldots ,\alpha_p\\\beta_1,\ldots 
,\beta_q\end{array}\Big\vert ~ x\right) :=\sum_{k\ge 0}\frac{(\alpha_1)_k
\cdots (\alpha_p)_kx^k}{(\beta_1)_k\cdots (\beta_q)_kk!}. \label{10.15}
\end{equation}
The symbol $(\alpha)_k$ is defined as follows:
\begin{equation}
(\alpha)_k=\frac{\Gamma (\alpha +k)}{\Gamma (\alpha )}=\alpha (\alpha +1)
\cdots (\alpha +k-1), \ \ k\ge 0. \label{10.16}
\end{equation}
In terms of hypergeometric series, formula (\ref{10.5}) takes the form

$\langle\psi ~|~\psi\rangle_{\rm reg}=$
\begin{equation}
=\left\{\prod_{l=-L}^L\frac{1}{M(l)+1}\right\}~_{2L+3}F_{2L+2}\left(
\begin{array}{c}3/2,1,-M(-L),\ldots ,-M(-L)\\ 1/2,M(-L)+2,\ldots ,M(L)+2
\end{array}\Big\vert ~1\right). \label{10.17}
\end{equation}
Using Dougall's formula [20] for the completely balanced series $_5F_4$ 
we obtain an especially simple expression for the square of the norm of 
the wave function for a periodic chain with three nodes ($L=1$):

$\langle\psi ~|~\psi\rangle_{\rm reg}=$
\begin{equation}
\frac{\Gamma (M(-1)+1)\Gamma (M(0)+1)\Gamma (M(1)+1)\Gamma 
(M(-1)+M(0)+M(1)+2)}{\Gamma (M(-1)+M(0)+2)\Gamma (M(-1)+M(1)+2)
\Gamma (M(0)+M(1)+2)}. \label{10.18}
\end{equation}

\vskip 0.5cm
{\bf \S 11. The multi point correlator.}
\vskip 0.5cm

In \S 8 it was shown that the correlator for the quantum Heisenberg model 
and the correlator for the modified classical Heisenberg model are 
connected by the relation
\begin{equation}
\left\langle\prod_{j=1}^N\wh S^{a_j}(r_j)\right\rangle_{kb}=\sum_{j=1}^N
(s(r_j)+1)\left\langle\prod_{j=1}^Nn^{a_j}(r_j)\right\rangle . \label{11.1}
\end{equation}
Here $r_N>r_{N-1}>\cdots >r_2>r_1$. It is clear that the correlator is 
equal to 0 for odd $N$. The right side of formula (\ref{11.1}) can be 
computed by means of commuting family of transfer matrices in analogy to 
the computation of the two--point correlator on the basis of 
(\ref{10.9}). The transfer matrix between two nearest nodes 
$n^{a_j}(r_j)$ and $n^{a_{j+1}}(r_{j+1})$ can be  by means of 
the polynomials $\Lambda^{A_{k_j}}_{k_j}$ (we recall that we call $k_j$ 
the spin of the tensor $\Lambda^{A_{k_j}}$). The difference of two 
nearest spins $k_j$ is equal to $k_{j+1}-k_j=\pm 1$ (see (\ref{9.10})). 
Thus, the spins of the tensors $\Lambda^A$ form a sequence 
$k_0,k_1,\ldots ,k_N$ of nonnegative integers such that
\begin{equation}
k_0=0, \ |k_{j+1}-k_j|=1, \ j=0,\ldots ,N-1, \ k_N=0. \label{11.2}
\end{equation}
We use the sequences (\ref{11.2}) to write out a formula for the 
multipoint correlator in explicit form. Since the function 
$\Lambda^A_{k_j-1}$ is multiplied by $n^{a_j}(r_j)$ at the vertex $r_j$, 
we must use the projections (\ref{9.11}). We denote by $X(k,M)$ the 
normalized eigenvalue of (\ref{9.8}):
\begin{equation}
X(k,M)=\frac{Z(k,M)}{Z(0,M)}=\frac{(-M)_k}{(M+2)_k}. \label{11.3}
\end{equation}
The symbol $(\alpha )_k$ is given by formula (\ref{10.16}).

We first write out the answer for correlation functions in the 
one  dimensional case (all the $M(l)$ are the same, $M(l)=M=s$; see 
(\ref{1.9})).
\begin{th} \label{t11.1} We have the equality
\begin{equation}
\left\langle\prod_{j=1}^Nn^{a_j}(r_j)\right\rangle =\sum_{\{ k\}}
\prod_{j=1}^{N-1}\{ X(k_j,M)\}^{r_{j+1}-r_j}P_0(n^{a_N}\cdots P_{k_j}
(n^{a_1a_2}))\cdots ). \label{11.4}
\end{equation}
\end{th}

The summation in (\ref{11.4}) goes over all possible sequences 
(\ref{11.2}). We remark that the last factor on the right side of 
(\ref{11.4}) is a $c$--number equal to a linear combination of products 
of the delta functions $\delta^{a_j}_{a_k}$ (and does not depend on the 
component $n^a$ of the vector ${\bf n}$ and the indices of the nodes 
$r_j$). It is easy to see that
$$X(1,M)=\frac{-M}{M+2}, \ \  X(2,M)=\frac{M(M-1)}{(M+2)(M+3)}.
$$

Using equalities (\ref{9.14}) for the projections, we obtain formulas 
(\ref{0.8}) and (\ref{0.9}) ($s=M$). We note that (\ref{11.4}) for even 
$N$ gives an answer for multipoint correlation functions also for an open 
chain (in the thermodynamic limit everything remains unchanged).

We now present an answer for correlation functions in the inhomogeneous 
case (all the $M(l)$ are distinct; see \S 2).
\begin{th} \label{t11.2} We have the equality
\begin{equation} 
\left\langle\prod_{j=1}^Nn^{a_j}(r_j)\right\rangle =\sum_{\{ k\}}
\prod_{j=1}^{N-1}\prod_{l=r_j}^{r_{j+1}-1}X(k_j,M(l))P_0(n^{a_N}\cdots
P_{k_3}(n^{a_3}P_{k_2}(n^{a_1a_2}))\cdots ). \label{11.5}
\end{equation}
\end{th}
The summation in (\ref{11.5}) goes over all possible sequences 
(\ref{11.2}).

The formula for the correlators (\ref{0.12}) for quasicrystals follows 
from (\ref{11.5}).

We make some remarks regarding the sequences (\ref{11.2}) and the 
projections (\ref{9.13}). First of all, it is natural to consider 
sequences (\ref{11.2}) of length $N+1$ with the boundary condition 
$k_N=l$ for fixed $l$. Let $a_{l,N}$ be the number of them. It is clear 
that $l\equiv N~(\mod 2)$. Using induction to $N$, it is possible to show 
that
\begin{equation}
a_{l,N}=\left(\begin{array}{c}(N-l)/2\\ N\end{array}\right) -
\left(\begin{array}{c} (N-l)/2-1\\ N\end{array}\right) . \label{11.6}
\end{equation}
The full number of sequences (\ref{11.2}) of length $N+1$ ($k_N$ is not 
fixed) is equal to
\begin{equation}
\sum_{l\le N}a_{l,N}=\left ( \begin{array}{c}{N}\\ {[N/2]}\end{array}\right ). 
\label{11.7}
\end{equation}
>From (\ref{11.6}) it is evident that $a_{l,N}$ is equal to the number of 
standard Young tableaux of the form $((N+l)/2,(N-l)/2)$ and is also equal 
to the multiplicity of degeneracy of the irreducible representation of 
the Lie algebra $o(3)$ of signature $(N+l)/2$ in the $N$th tensor power 
of the fundamental representation. In our special case $l=0$, and hence
\begin{equation}
a_{0,N}=\left\{\begin{array}{ll}0 & {\rm if} \ N\equiv 1~(\mod 2)\\
\ds\frac{(2k)!}{k!(k+1)!} & {\rm if} \ N=2k, \ k \ {\rm an~integer}.
\end{array}\right. \label{11.8}
\end{equation}
We shall construct a bijection between the set of sequences $(k_0,\ldots 
,k_N)$ such that $k_0=0$, $k_N=l$, $k_j\in{\bf Z_+}$, $|k_i-k_{i+1}|=1$, 
$i=0,1,\ldots ,N-1$, and the set of standard Young tableaux $T$ of the 
form $\lambda =((N+l)/2, (N-l)/2)$. To this end we consider the sequence 
$k_0,k_1,\ldots ,k_N$. We write the number $i\in [1,N]$ in the first row 
of the diagram $\lambda$ if $k_{i-1}-k_i=-1$ and in the second row 
otherwise. The bijection has been constructed. We present an example. 
Suppose $\{ k_i\}=(0,1,2,3,2,1,0,1,0,1)$. Then
$$T=\begin{array}{ccccc} 1&2&3&7&9\\ 4&5&6&8\end{array} .
$$

We have thus obtained a representation for the $N$--point correlator 
(\ref{11.5}) in the form of a sum over standard Young tableaux of the 
form $(N/2,N/2)$. It would be desirable to have a more explicit formula 
for projection (\ref{9.13}), proceeding directly from the standard table 
$T$ of the form $((N+l)/2$,\break $(N-l)/2)$. This problem is in the stage of 
solution. We only make several remarks. We denote by $P_T$ the projection 
(\ref{9.13}) corresponding to the table $T$. It is clear that $P_T$ is  a 
linear combination of the polynomials $\Lambda^A_l$. For example, if
$$T=\begin{array}{ccccccc}1&3&\ldots &N-+1&N-l+1&\ldots &N\\
2&4&\ldots &N-l\end{array}
$$
then
\begin{equation}
P_T=\left(\frac{1}{2\lambda +2}\right)^{(N-l)/2}\Lambda_l^{\{N-l+1,\ldots 
,N\}}\prod_{j=1}^{(N-1)/2}\delta_{a_{2j}}^{a_{2j-1}}, \label{11.9}
\end{equation}
if
$$T=\begin{array}{cc}1&2\\ 3&4\end{array} ,
$$
then
\begin{equation}
P_T=\frac{1}{(2\lambda +2)(2\lambda 
+4)}\left\{\delta^{a_1}_{a_4}\delta^{a_2}_{a_3}+
\delta^{a_1}_{a_3}\delta^{a_2}_{a_4}-\frac{1}{\lambda +1}\delta^{a_1}_{a_2}
\delta^{a_3}_{a_4}\right\} , \label{11.10}
\end{equation}
if
$$T=\begin{array}{ccc} 1&2&4\\ 3\end{array} ,
$$
then
$$P_T=\frac{1}{2\lambda +4}\left\{\Lambda^{a_1a_4}\delta^{a_2}_{a_3}+
\Lambda^{a_2a_4}\delta^{a_1}_{a_3}-\frac{1}{\lambda +1}\delta^{a_1}_{a_2}
\delta^{a_3}_{a_4}\right\} .
$$

For the Lie algebra $o(3)$ in the preceding formulas it is necessary to 
set $\lambda =1/2$.

\vskip 0.5cm
{\bf \S 12. The $o(m)$--generalization of the modified classical 
Heisenberg model.}
\vskip 0.5cm

In the preceding sections we considered the $o(3)$--modified classical 
Heisenberg model. There is a natural generalization to the case of the 
Lie algebra $o(m)$ and irreducible representations corresponding to 
one--sided Young diagrams. The statistical sum for one--dimensional, 
classical Heisenberg model has the form
\begin{equation}
\Phi =\int_{S^{m-1}}\prod^L_{l=-L}d\Omega_l\prod_{l=-L}^{L-1}
\left(\frac{1-{\bf n}(l)\cdot{\bf n}(l+1)}{2}\right)^{M(l)}. \label{12.1}
\end{equation}

In (\ref{12.1}) the vector ${\bf n}\in S^{m-1}$, and the invariant 
measure $d\Omega$ is presented in Appendix B. We emphasize that now $\bf 
n$ is an $m$--component unit vector.

The integral operator $\wh K_M$ is defined in analogy to (\ref{9.1}):
\begin{equation}
(\wh K_Mf)({\bf n}_2)=\int_{S^{m-1}}d\Omega_l\left(\frac{1-{\bf 
n}_1\cdot{\bf n}_2}{2}\right)^Mf({\bf n}_1). \label{12.2}
\end{equation}
The eigenfunctions of the operator $\wh K_M$ are also given by formula 
(\ref{9.5}) in which the coefficient $\gamma_k(N)$ are
\begin{equation}
\gamma_k(N)=\prod_{j=1}^k\frac{-1}{m+2(N-j-1)}, \ \ \gamma_0(N)=1. 
\label{12.3}
\end{equation}
The tensor $\Lambda^{A_N}$ (see (\ref{9.5}), (\ref{12.3})) is 
characterized uniquely by the conditions of symmetry and tracelessness, 
and its principal part $n^{A_N}$. We note that the dimension of the 
tensor $\Lambda^{A_N}$ (for fixed $N$) is
\begin{eqnarray}
\left(\begin{array}{c}N+m-1\\ m-1\end{array}\right)&-&\left(\begin{array}{c}
N+m-3\\ m-1\end{array}\right) =\frac{2N+m-2}{m-3}\left(\begin{array}{c}
N+m-3\\ m-3\end{array}\right) \nonumber \\
&=&\left(\begin{array}{c}N+m-2\\ m-2\end{array}\right) 
+\left(\begin{array}{c}N+m-3\\ m-2\end{array}\right), \label{12.4}
\end{eqnarray}
and coincides with the dimension of the irreducible representation of the 
Lie algebra $o(m)$ with highest weight $(N,0,\ldots ,0)$. We shall 
formulate the basic properties of the polynomials $\Lambda^{A_N}$ as a 
theorem (in which $\lambda =m/2-1$).

We consider the polynomials
\begin{equation}
\wt\Lambda^{A_N}({\bf n})=\sum_{l\ge 0}\frac{r^{2l}}{2^l(-\lambda -N+1)_l}
\sum_{A_{2l}\subset A_N}n^{A_N/A_{2l}}\delta (A_{2l}). \label{12.5}
\end{equation}
Here ${\bf n}\in{\bf R}^m$, $r^2=\ds\sum^m_{a=1}(n^a)^2$. Hence, 
$\Lambda^{A_N}=\wt\Lambda^{A_N}|_{S^{m-1}}$.
\begin{th} \label{t12.1} 1. The tensor $\wt\Lambda^{A_N}$ is traceless; 
in other words, if $A_N=$\break $\{ a_1,a_2,\ldots \}$, then
\begin{equation}
\tr_{\{ 
a_1,a_2\}}\wt\Lambda^{A_N}:=\sum_{a_1=1}^m\sum_{a_2=1}^m\wt\Lambda^{A_N}
({\bf n})\delta^{a_1}_{a_2}=0. \label{12.6}
\end{equation}

2. Let $X$ be a symmetric polynomial in the variables $n^1,\ldots ,n^m$, 
and suppose $\Pi_{\{ a_1,a_2\}}x=0$. We suppose that 
$X^{\max}=\ds\sum_{A_N}\alpha_{A_N}\cdot n^{A_N}$. Then\break 
$X=\ds\sum_{A_N}\alpha_{A_N}\cdot\wt\Lambda^{A_N}({\bf n})$.

3. $\Delta\wt\Lambda^{A_N}\equiv 0$, where $\Delta$ is the Laplace 
operator for ${\bf R}^m$. Hence, $\wt\Lambda^{A_N}$ is a harmonic 
polynomial of degree $N$, and the polynomials $\wt\Lambda^{A_N}$, 
$A_N\subset [1,\ldots ,m]^N$ with the condition $\#\{ a_i\in A_N:a_i=1\}\le 
1$, form a basis in the space of harmonic polynomials of degree $N$ in 
${\bf R}^m$.

4. Let $\Delta_S$ be the Laplace operator on the sphere $S^{m-1}$. Then 
$A_S\Lambda^{A_N}=-N(N+m-2)$.

5. We consider the operators
\begin{equation}
E_{jk} =n^k\frac{\partial}{\partial n^j}-n^j\frac{\partial}{\partial 
n^k}, \ \ 1\le j\ne k\le m. \label{12.7}
\end{equation}
The generators $E_{jk}$ generate the Lie algebra $SO(m)$:
\begin{equation}
[E_{jk},E_{pq}]=\delta_{kp}E_{jq}, \ \ {\rm if} \ j<k, \ p<q. \label{12.8}
\end{equation}
The action of the generators $E_{jk}$ on the polynomials $\Lambda^{A_N}$ 
(for fixed $N$) is given by the formula
\begin{equation}
E_{jk}\Lambda^{A_N}=\sum_{a\in A_N}\{\Lambda^{(A_N\cup\{ k\})\setminus\{ 
a\}}\delta_j^a-\Lambda^{(A_N\cup\{ j\})\setminus\{ a\}}\delta_k^a\}. 
\label{12.9}
\end{equation}

6. For each polynomial $f({\bf n})$ of degree $N$, ${\bf n}\in{\bf R}^m$, 
we define its harmonic projection (see [19]; here $\lambda =m/2-1$):
\begin{equation}
(Hf)({\bf n})=\sum_{l=0}^{[N/2]}\frac{r^{2l}(\Delta^lf)({\bf n})}
{2^{2l}l!(-\lambda -N+1)_l}. \label{12.10}
\end{equation}
Then the polynomial $\wt\Lambda^{A_N}$ is the harmonic projection of the 
monomial $n^{A_N}$:
$$\wt\Lambda^{A_N}({\bf n})=(Hn^{A_N})(n).
$$
Moreover, it is easy to see that
\begin{equation}
\Delta^kn^{A_N}=2^k\cdot k!\sum_{A_{2k}\subset A_N}n^{A_N\setminus A_{2k}}
\delta (A_{2k}). \label{12.11}
\end{equation}

7. Let $A_N=\{\underbrace{a,\ldots ,a}_N\}$. Then
\begin{equation}
\wt\Lambda^{A_N}({\bf n})=\frac{N!\Gamma (\lambda )r^N}{2^N\Gamma 
(\lambda +N)}\cdot C_N^{\lambda}\left(\frac{n^a}{r}\right) .\label{12.12}
\end{equation}
Here the $C_N^{\lambda}$ are the Gegenbauer polynomials whose definition 
is presented in Appendix B.
\end{th}

It is interesting to note that formula (\ref{9.5}) for $\Lambda^{A_N}$ 
can be inverted, and the monomials $n^{A_N}$ (see (\ref{9.3}) can be 
expressed in terms of the harmonic polynomials $\wt\Lambda^{A_N}$:
\begin{equation}
n^{A_N}=\sum_{l=0}^{[N/2]}r^{2l}\beta_l(N)\sum_{A_{2l}\subset A_N}
\wt\Lambda^{A_N\setminus A_{2l}}\delta (A_{2l}). \label{12.13}
\end{equation}
Here $\beta_l(N)=\gamma_l(M-l+1)$, where
\begin{equation}
\gamma_l(M)=\frac{1}{2^l(-\lambda -M+1)_l}. \label{12.14}
\end{equation}

The identity (\ref{12.13}) plays an important role in further 
considerations, since it makes it possible to solve the problem of 
decompositing the tensor product of two irreducible (single--sided) 
representations of the lie algebra $o(m)$ into irreducible components. In 
particular,
\begin{eqnarray}
\Lambda_N^{A_N}\cdot n^{a_{N+1}}&=&\Lambda_{N+1}^{A_N\cup a_{N+1}}+
\frac{1}{2(\lambda +N)}\sum_{j=1}^N\Lambda_{N-1}^{a_1\cdots\hat a_j\cdots
a_N}\delta^{a_j}_{a_{N+1}} \label{12.15}\\
&-&\frac{1}{2(\lambda +N)(\lambda +N-1)}\sum_{1\le i<j\le N}
\Lambda_{N-1}^{a_1\cdots \hat a_i\cdots \hat a_j\cdots a_{N+1}}
\delta^{a_j}_{a_i}.\nonumber 
\end{eqnarray}
We thus consider projections analogous to (\ref{9.11}) and (\ref{9.12}):
\begin{equation}
P_{N+1}(\Lambda_N^{A_N}\cdot n^{a_{N+1}})=\Lambda_{N+1}^{A_N\cup 
a_{N+1}}, \label{12.16}
\end{equation}
\begin{equation}
P_{N-1}(\Lambda_N^{A_N}\cdot n^{a_{N+1}})=\Lambda_N^{A_N}\cdot n^{a_{N+1}}
-\Lambda_{N+1}^{A_N\cup a_{N+1}}, \label{12.17}
\end{equation}
$$P_k(\Lambda_N^{A_N}\cdot n^{a_{N+1}})=0, \ \ {\rm if} \ k\ne N\pm 1.
$$

We note that the operators $\wh K_M$, as in the case $m=3$, form a 
commuting family: 
\begin{equation}
[\wh K_{M_1},\wh K_{M_2}]=0. \label{12.18}
\end{equation} 
It follows from Theorem \ref{12.1} that the polynomials 
$\Lambda^{A_N}$ are eigenfunctions of the operators $\wh K_M$: 
\begin{equation} 
\wh K_M\Lambda_N^{A_N}=Z_m(N,M)\Lambda_N^{A_N}. \label{12.19} 
\end{equation} 

The eigenvalue $Z_m(N,M)$ can be computed from the formula
\begin{equation}
Z_m(N,M)=\frac{\Gamma (2\lambda +1)}{\Gamma (\lambda +\frac{1}{2})}
\frac{\Gamma (M+\lambda +\frac{1}{2})(-M)_N}{\Gamma (M+2\lambda +1)
(M+2\lambda +1)_N}. \label{12.20}
\end{equation}
We have used here the notation (\ref{10.16}). 
The proof of formulas (\ref{12.19}) and (\ref{12.20}) is  presented in 
Appendix B. Another important characteristic is
\begin{equation}
X_m(k,M)=\frac{Z_m(k,M)}{Z_m(0,M)} \label{12.21}
\end{equation}
which can be computed by means of (\ref{12.20}) and is equal to
\begin{equation}
X_m(k,M)=\frac{(-M)_k}{(M+2\lambda +1)_k}. \label{12.22}
\end{equation}
We can now conclude the computation of the multipoint correlation 
functions for the one--dimensional, modified, classical Heisenberg 
$O(m)$--model. The answer is given by formulas (\ref{11.4}) and 
{\ref{11.5}) in which the $X(k_j,M)$ are now replaced by (\ref{12.22}), 
the projections (\ref{9.11}) and (\ref{9.12}) are replaced by the 
projections (\ref{12.16}) and (\ref{12.17}), and the summation both in 
(\ref{11.4}) and (\ref{11.5}) goes over sequences $(k_0,k_1,\ldots ,k_N)$ 
such that $k_0=k_n=0$, $k_i\in{\bf Z}_+$, $|k_i-k_{i+1}|=1$, 
$i=0,1,\ldots ,N-1$.

In conclusion we present formulas for square of the norm of the VBS wave 
function of the classical Heisenberg $O(m)$--model for an open and 
periodic one--dimensional chain (we recall that $\lambda =m/2-1$):
\begin{equation}
\langle\psi ~|~\psi\rangle_0 =\left\{\frac{\Gamma (2\lambda +1)}
{\Gamma (\lambda +\frac{1}{2})}\right\}^{2L}\prod_{l=-L}^{L-1}
\frac{\Gamma (M(l)+\lambda +\frac{1}{2})}{\Gamma (M(l)+2\lambda +1)}. 
\label{12.23}
\end{equation}
We note that if all $M(l)\in{\bf Z}_+$, then
\begin{equation}
\langle\psi ~|~\psi\rangle_0=\prod_{l=-L}^{L-1}\frac{(\lambda 
+\frac{1}{2})_{M(l)}}{(2\lambda +1)_{M(l)}}. \label{12.24}
\end{equation}
Formulas (\ref{12.23}), (\ref{12.24}) for $\lambda =1/2$ go over into 
(\ref{10.2}):
\begin{eqnarray}
\langle\psi ~|~\psi\rangle_{\rm reg}&=&\prod_{l=-L}^L\frac{\Gamma 
(2\lambda +1)\Gamma (M(l)+\lambda +\frac{1}{2})}{\Gamma (\lambda 
+\frac{1}{2})\Gamma (M(l)+2\lambda +1)} \label{12.25}\\
&\times&_{2L+3}F_{2L+2}\left(\begin{array}{l}2\lambda , \lambda +1, 
-M(-L),\ldots ,-M(L)\\ \lambda ,M(-L)+2\lambda +1,\ldots , M(L)+2\lambda 
+1 \end{array}\Big\vert ~1\right) .\nonumber
\end{eqnarray}

\vskip 0.5cm
{\bf \S 13. An algebraic approach to the computation of correlators for 
the multidimensional Heisenberg model.}
\vskip 0.5cm

We consider the statistical sum for the classical Heisenberg 
$O(m)$--model on a multidimensional lattice
\begin{equation}
\Phi (M)=\int_{S^{m-1}}\prod_{\rm sites}d\Omega\prod_{\langle ij\rangle}
\left(\frac{1-{\bf n}_i{\bf n}_j}{2}\right)^M. \label{13.1}
\end{equation}
It is clear that $\Phi (M)$ is equal to the square of the norm of the VBS 
wave function (\ref{2.5}) for quantum Heisenberg model describing the 
interaction of spins $s=Mz/2$ (here $z$ is the coordination number of 
the lattice). Computation of correlation functions of the quantities 
$(1-{\bf n}_k\cdot{\bf n}_l)/2$, where $k,l$ are nearest neighbors, is of 
major interest. First of all the mean value of $(1-{\bf n}_1\cdot{\bf 
n}_2)/2$ is the N\'eel order parameter
\begin{equation}
\left\langle\frac{1-{\bf n}_1{\bf n}_2}{2}\right\rangle =\frac{\int\prod_j
d\Omega\prod_{\langle ij\rangle}((1-{\bf n}_i\cdot{\bf n}_j)/2)^M
((1-{\bf n}_1\cdot{\bf n}_2)/2)}{\int\prod_jd\Omega\prod_{\langle 
ij\rangle}((1-{\bf n}_i\cdot{\bf n}_j)/2)^M}. \label{13.2}
\end{equation}
Equality to one of the mean of $(1-{\bf n}_1\cdot{\bf n}_2)/2$ is 
equivalent to the existence of a N\'eel order in the system. For 
one--dimensional lattice of $O(m)$--spins $s$ the N\'eel order parameter 
is computed in (\ref{12.24}):
\begin{equation}
\left\langle\frac{1-{\bf n}_1\cdot{\bf n}_2}{2}\right\rangle =
\frac{s+\lambda +\frac{1}{2}}{s+2\lambda +1}<1. \label{13.3}
\end{equation}
Here $\lambda =m/2-1$. Equality (\ref{13.3}) shows the absence of N\'eel 
order for the model considered. It is interesting to note that for 
one--dimensional Heisenberg chain there is no correlation between N\'eel 
order parameters:
$$\langle (1-{\bf n}(1)\cdot{\bf n}(2))(1-{\bf n}(j)\cdot{\bf 
n}(j+1))\rangle =\langle 1-{\bf n}(1)\cdot{\bf n}(2)\rangle\langle 1-{\bf 
n}(j)\cdot{\bf n}(j+1)\rangle .
$$

To compute the correlation functions of several quantities $(1-{bf 
n}_k\cdot{\bf n}_l)/2$ we consider the generating function for the 
correlators
\begin{equation}
\Phi_{\Gamma}(\{ M_{ij}\})=\int_{S^{m-1}}\prod_{\rm sites}d\Omega
\prod_{\langle ij\rangle}\left(\frac{1-{\bf n}_i\cdot{\bf 
n}_j}{2}\right)^{M_{ij}}. \label{13.4}
\end{equation}
In the general case all the alternating numbers $M_{ij}$ are assumed to 
be distinct. It is most natural to consider the Heisenberg model for the 
complete graph (all the vertices are joined by edges) with $p+1$ vertices
\begin{equation}
\Phi_{p+1}(\{ M_{ij}\})=\int_{S^{m-1}}\prod_{j=1}^{p+1}d\Omega
\prod_{1\le i<j\le p+1}\left(\frac{1-{\bf n}_i\cdot{\bf n}_j}{2}
\right)^{M_{ij}}. \label{13.5}
\end{equation}
We emphasize again that we consider the alternating numbers $M_{ij}$ as 
independent parameters. In this section we shall prove that 
$\Phi_{\Gamma}$ is a rational function of (integer) parameters $M_{ij}$. 
We note that an arbitrary graph can be obtained from complete graph by 
annihilating some of the parameters $M_{ij}$. Indeed, the condition 
$M_{ij}=0$ is equivalent to the absence of the edge $\langle ij\rangle$ 
in the graph.

We proceed to the investigation of the statistical sum (\ref{13.5}) for 
the Heisenberg $O(m)$--model on a complete graph with $p+1$ vertices. We 
wish to find recurrence relations between the functions $\Phi$. For this 
we first compute the integral with respect to a fixed invariant measure 
$d\Omega_j$ of the integral (\ref{13.5}) in terms of generalized 
hypergeometric functions. We must thus compute the integral
\begin{equation}
\int_{S^{m-1}}d\Omega_0\prod_{j=1}^p\left(\frac{1-{\bf n}_0\cdot{\bf n}_j}
{2}\right)^{M_{0j}}. \label{13.6}
\end{equation}
Before presenting an explicit expression for the integral (\ref{13.6}), 
we give some definitions and examples. We consider collections of complex 
numbers $z_{ij}$ and nonnegative integers $k_{ij}$ ($1\le i<j\le p$). We 
denote these collections by $z^{(p)}$ and $k^{(p)}$. For the collection 
$k^{(p)}$ we define
$$|k^{(p)}|:=\sum_{1\le i<j\le p}k_{ij}, \ \ 
(k^{(p)})!:=\prod_{1\le i<j\le p}(k_{ij})!, 
$$
\begin{equation} 
\label{13.7}
\end{equation}
$$k_i^{(p)}=\sum_{1\le l<i}k_{li}+\sum_{i\le l\le p}k_{il}.
$$
We now define the generalized hypergeometric series
$$F^{(p)}\left(\begin{array}{l}\alpha_1,\ldots ,\alpha_p\\\beta\end{array}
\Big\vert ~z^{(p)}\right)
$$
depending on $p(p-1)/2$ complex variables $z^{(p)}$ and on $p+1$ real 
parameters $\alpha_1,\ldots ,\alpha_p,\beta$. The function $F^{(p)}$ is 
given by means of $\frac{1}{2}p(p-1)$--fold sum over the collections 
$k^{(p)}$:
\begin{equation}
F^{(p)}\left(\begin{array}{l}\alpha_1,\ldots ,\alpha_p\\\beta\end{array}
\Big\vert ~z^{(p)}\right):=\sum_{k^{(p)}}
\frac{\ds\prod^p_{i=1}(\alpha_i)k_i^p}{(\beta )_{|k|^{(p)}}}
\prod_{1\le i<j\le p}\frac{z_{ij}^{k_{ij}}}{(k_{ij})!}. \label{13.8}
\end{equation}
Here we use the notation (\ref{10.16}):
$$(\alpha )_k=\frac{\Gamma (\alpha +k)}{\Gamma (\alpha )}.
$$

We note that if the $\alpha_j$ are nonnegative integers, then the series 
(\ref{13.8}) breaks off and the function
$$F^{(p)}\left(\begin{array}{l}\alpha_1,\ldots ,\alpha_p\\\beta\end{array}
\Big\vert ~z^{(p)}\right)
$$
can be represented as a finite sum. We present examples of the functions 
$F^{(p)}$ for $p=2,3$ ($F^{(1)}\equiv 1$):
\begin{equation}
F^{(2)}\left(\begin{array}{l}\alpha_1,\alpha_2\\\beta\end{array}\Big\vert 
~z\right) =\sum_{k\ge 0}\frac{(\alpha_1)_k(\alpha_2)_k}{(\beta )_k}\cdot
\frac{z^k}{k!}. \label{13.9}
\end{equation}
\begin{eqnarray}
F^{(3)}\left(\begin{array}{l}\alpha_1,\alpha_2,\alpha_3\\\beta\end{array}
\Big\vert ~z_1,z_2,z_3\right)& =&\sum\frac{(\alpha_1)_{k_2+k_3}
(\alpha_2)_{k_1+k_3}(\alpha_3)_{k_1+k_2}}{(\beta 
)_{k_1+k_2+k_3}}\nonumber\\
&\times&\frac{z_1^{k_1}}{k_1!}\cdot\frac{z_2^{k_2}}{k_2!}\cdot
\frac{z_3^{k_3}}{k_3!}. \label{13.10}
\end{eqnarray}
We now formulate the result of evaluating the integral (\ref{13.6}).
\begin{th} \label{t13.1} Let ${\bf n}_0,{\bf n}_1,\ldots ,{\bf n}_p\in 
S^{m-1}$. Then
$$\int_{S^{m-1}}d\Omega\prod_{j=1}^p\left(\frac{1-{\bf n_0\cdot n_j}}{2}
\right)^{M_{0j}}=\frac{\Gamma (2\lambda +1)}{\Gamma (\Lambda +\frac{1}{2})}
\cdot\frac{\Gamma (M_{01}+\cdots +M_{0p}+\lambda +\frac{1}{2})}
{\Gamma (M_{01}+\cdots +M_{0p}+2\lambda +1)}
$$
\begin{equation}
\times F^{(p)}\left(\begin{array}{l}-M_{01},-M_{02},\ldots ,-M_{0p}\\
-M_{01}-M_{02}-\cdots -M_{0p}-\lambda +\frac{1}{2}\end{array}\Big\vert ~
\left\{\ds\frac{1-{\bf n_i\cdot n_j}}{2}\right\}_{1\le i<j\le p}\right) . 
\label{13.11}
\end{equation}
\end{th}

We recall that $\lambda =m/2-1$. We again emphasize that if all the 
$M_{0j}\in{\bf Z}_+$, then the series on the right side of (\ref{13.11}) 
contains only a finite number of terms.

>From Theorem \ref{t13.1} we obtain a recurrence relation for the 
statistical sum (\ref{13.5}) of the classical Heisenberg $o(m)$--model on 
a complete graph. We denote the vertices of the complete graph with $p+1$ 
vertices by $l=0,1,\ldots ,p$. We decompose the set of alternating 
numbers $\{ M_{lm}\}=\{ M_{oj}\}\cup\{ M_{ij}\}$ where $1\le i,j\le p$. 
We consider the complete graph with $p$ vertices as a subgraph of the 
complete graph with $p+1$ vertices for which the vertex $l=0$ and the 
edges $\langle 0,1\rangle ,\ldots ,\langle 0,p\rangle$ have been removed.
\begin{th} \label{t13.2} The statistical sums $\Phi_p$ and $\Phi_{p+1}$ 
are connected by the relation
$$\Phi_{p+1}(\{ M_{0j}\}\cup\{ M_{ij}\})=\frac{\Gamma (2\lambda +1)}
{\Gamma (\lambda +\frac{1}{2})}\cdot\frac{\Gamma (M_{01}+\cdots 
+M_{0p}+\lambda +\frac{1}{2})}{\Gamma (M_{01}+\cdots +M_{0p}+2\lambda +1)}
$$
\begin{equation}
\times\sum_{k^{(p)}}\frac{\ds\prod_{j=1}^p(-M_{0j})_{k_j^{(p)}}}
{(k^{(p)})!(-\ds\sum_{j=1}^pM_{oj}-\lambda +\frac{1}{2})_{|k^{(p)}|}}\cdot
\Phi_p(\{ M_{ij}+k_{ij}\}). \label{13.12}
\end{equation}
\end{th}

The proof follows immediately from Theorem \ref{t13.1} and formula 
(\ref{13.8}). As an example we present the formulas for $\Phi_p$ for 
$p\le 3$:
\begin{equation}
\Phi_0=\Phi_1=1, \ \ \Phi_2(M)=\frac{\Gamma (2\lambda +1)}{\Gamma 
(\lambda +\frac{1}{2})}\cdot\frac{\Gamma (M+\lambda +\frac{1}{2})}
{\Gamma (M+2\lambda +1)}, \label{13.13}
\end{equation}
\begin{eqnarray}
\Phi_3(M_1,M_2,M_3)&=&\frac{[\Gamma (2\lambda +1)]^2\Gamma 
(M_1+M_2+M_3+2\lambda +1)}{[\Gamma (\lambda +\frac{1}{2})]^3} \nonumber \\
&\times&\prod_{j=1}^3\frac{\Gamma (M_j+\lambda +\frac{1}{2})}
{\Gamma (M_1+M_2+M_3-M_j+2\lambda +1)}. \nonumber
\end{eqnarray}
It should be noted that Theorem \ref{t13.1} can be used not only for 
complete graphs. We consider, for example, a graph $\Gamma$ containing a 
vertex $l$ of multiplicity 2. It may be assumed that $l=1$ and the vertex 
1 is joined only with vertices 2 and 3. We denote by $\wh\Gamma$ the 
graph obtained from $\Gamma$ by removing vertex 1 and the edges $\langle 
12\rangle$ and $\langle 13\rangle$. The statistical sums corresponding to 
the graphs $\Gamma$ and $\wh\Gamma$ are connected by relation
\begin{eqnarray}
\Phi_{\Gamma}(M_{12},M_{13},M_{23},\ldots )&=&\frac{\Gamma (2\lambda +1)}
{\Gamma (\Lambda +\frac{1}{2})}\cdot\frac{\Gamma (M_{12}+M_{13}+\lambda 
+\frac{1}{2})}{\Gamma (M_{12}+M_{13}+2\lambda +1)} \label{13.14}\\
&& \nonumber \\
&\times&\sum_{k\ge 0}\frac{(-M_{12})_k(-M_{13})_k}{k!(-M_{12}-M_{13}-
\lambda +\frac{1}{2})_k}\Phi_{\wh\Gamma}(M_{23}+k,\ldots ). \nonumber
\end{eqnarray}
The statistical sum $\Phi_{\Gamma}$ was computed in \S\S 10, 12 for a 
periodic, one--dimen\-sio\-nal chain (a polygon). If (\ref{12.25}) is 
substituted into (\ref{13.14}) we obtain a new identity for generalized 
hypergeometric series:
\begin{eqnarray}
&&_{p+2}F_{p+1}\left(\begin{array}{l}2\lambda ,\lambda +1, -m_1,\ldots , 
-m_p\\
\lambda ,m_1+2\lambda +1, \ldots ,m_p+2\lambda +1\end{array}\Big\vert 
~1\right) \label{13.15} \\
&& \nonumber\\
&=&\frac{\Gamma (\lambda +\frac{1}{2})}{\Gamma (2\lambda +1)}\cdot
\frac{\Gamma (m_{p-1}+2\lambda +1)\Gamma (m_p+2\lambda +1)\Gamma 
(m_{p-1}+m_p+\lambda +\frac{1}{2})}{\Gamma (m_{p-1}+\lambda +\frac{1}{2})
\Gamma (m_p+\lambda +\frac{1}{2})\Gamma (m_{p-1}+m_p+2\lambda +1)} 
\nonumber \\
&& \nonumber\\
&\times&\sum_{k\ge 0}\frac{(-m_{p-1})_k(-m_p)_k(\lambda +\frac{1}{2})_k}
{k!(-m_{p-1}-m_p-\lambda +\frac{1}{2})_k(2\lambda +1)_k} \nonumber \\
&& \nonumber\\
&\times&_{p+1}F_p\left(\begin{array}{l}2\lambda ,\lambda +1, -m_1,\ldots , 
-m_{p-2},-k\\
\lambda ,m_1+2\lambda +1, \ldots ,m_{p-2}+2\lambda +1,k+2\lambda +1
\end{array}\Big\vert ~1\right) .\nonumber
\end{eqnarray}
In particular, for $p=3$ we arrive at Doudall's formula [20] for the 
completely balanced series $_5F_4$. For $p=4$ we arrive at the formula 
for the transformation of the completely balanced series $_6F_5$ into the 
Saalsch\"tz series $_4F_3$:

$_6F_5\ds\left(\begin{array}{l}2\lambda ,\lambda +1,-m_1,-m_2,-m_3,-m_4\\
\lambda ,m_1+2\lambda +1,\ldots ,m_4+2\lambda +1\end{array}\Big\vert ~1
\right)$
\begin{eqnarray}
&=&\frac{\Gamma (\lambda +\frac{1}{2})\Gamma (m_1+2\lambda +1)\Gamma (m_2+
2\lambda +1)\Gamma (m_1+m_2+\lambda +\frac{1}{2})}{\Gamma (2\lambda +1)
\Gamma (m_1+\lambda +\frac{1}{2})\Gamma (m_2+\lambda +\frac{1}{2})\Gamma
(m_1+m_2+\lambda +1)} \nonumber \\
&& \nonumber\\
&\times& _4F_3\left(\begin{array}{l}-m_1,m_2,m_3+m_4+2\lambda +1,\lambda 
+\frac{1}{2}\\-m_1-m_2-\lambda +\frac{1}{2},m_3+2\lambda +1,m_4+2\lambda 
+1\end{array}\Big\vert ~1\right) . \label{13.16}
\end{eqnarray}
We note that the recurrence relation (\ref{13.14}) makes it possible to 
compute the statistical sum $\Phi_{\Gamma}$ for the following graph 
$\Gamma$:
$$\Gamma 
=\left\{\begin{array}{ccc}&m_1\\m_4
&\hbox{\setlength{\unitlength}{0.3cm}
\begin{picture}(1.5,2.5)(0.7,1.25)
\put(0,0){\line(1,0){2}}
\put(0,3){\line(0,-1){3}}
\put(0,3){\line(1,0){2}}
\put(2,0){\line(0,1){3}}
\put(2,0){\line(-2,3){2}}
\put(0.1,0.4){\hbox{${\scriptstyle m}_5$}}
\end{picture}}&
m_2\\ \\ &m_3\end{array}\right\} .
$$

$$\Phi_{\Gamma}=\frac{[\Gamma (2\lambda +1)]^3}{[\Gamma (\Lambda 
+\frac{1}{2})]^4}\frac{\Gamma (m_1+m_2+\lambda +\frac{1}{2})\Gamma 
(m_3+\lambda +\frac{1}{2})\Gamma (m_4+\lambda +\frac{1}{2})}
{\Gamma (m_1+m_2+2\lambda +1)\Gamma (m_3+m_4+2\lambda +1)} 
$$

$\times ~\ds\frac{\Gamma (m_5+\lambda +\frac{1}{2})\Gamma 
(m_3+m_4+m_5+2\lambda +1)}{\Gamma (m_3+m_5+2\lambda +1)
\Gamma (m_4+m_5+2\lambda +1)}$ 
\vskip 0.3cm

$\times ~_4F_3\left(\begin{array}{l}-m_1,m_2,m_3+m_4+m_5+2\lambda +1,m_5+\lambda 
+\frac{1}{2}\\-m_1-m_2-\lambda +\frac{1}{2},m_3+m_5+2\lambda +1,m_4+m_5
+2\lambda +1\end{array}\Big\vert ~1\right) .$ 
\vskip 0.3cm

The proof of Theorem \ref{t13.1} makes use of a large number of different 
identities for sums of products of binomial coefficients; they are rather 
cumbersome and will be presented in a separate paper. Here we note only 
that Theorem \ref{t13.1} follows from the more general Theorem B, whose 
formulation is presented in Appendix B.

\vskip 0.5cm
{\bf Appendix A.}
\vskip 0.5cm

We represent the operator $a^+_l$ as the operator of multiplication by 
$x_l$ and the operator $b^+_l$ as the operator of multiplication by 
$y_l$. The spin operators at the $l$th node can then be represented in 
the form
$$S_l^+=x_l\frac{\partial}{\partial y_l}; \ \ 
S^-_l=\frac{\partial}{\partial x_l};\eqno  (A.1)
$$
$$2S^z_l=x_l\frac{\partial}{\partial x_l}-y_l\frac{\partial}{\partial y_l}.
$$
We consider the space of the irreducible representation $V_{s_l}$ of spin 
$s$. A basis can be taken in the form
$$v_{\mu}=x_l^{s_l+\mu}y_l^{s_l-\mu}; \ \mu =-s_l,s_l+1,\ldots ,s_l-1,s_l.
\eqno (A.2)
$$
The spin operators act on the basis in the following manner:
$$S^+_lv_{\mu}=(s_l-\mu )v_{\mu +1};
$$
$$S^-_lv_{\mu}=(s_l+\mu )v_{\mu -1};\eqno (A.3)
$$
$$S^z_lv_{\mu}=\mu v_{\mu}.
$$
We now consider the tensor product of two irreducible representations 
$V_{s_l}\otimes V_{s_k}$. The spin operators add:
$$S^{\pm}=S^{\pm}_l+S^{\pm}_k, \ \ S^z=S^z_l+S^z_k. \eqno (A.4)
$$
The tensor product decomposes into a direct sum of irreducible 
representations:
$$ V_{s_l}\otimes V_{s_k}=\sum_{J=|s_l+s_k|}^{s_l+s_k}V_J. \eqno (A.5)
$$
We shall construct the leading vector $v_J$ of the irreducible 
representation $V_J$. We seek it in the form
$$v_J=\sum_{\mu_j+\mu_k=J}a_{\mu_l,\mu_k}x_l^{s_l+\mu_l}y_l^{s_l-\mu_l}
x_k^{s_k+\mu_k}y_k^{s_k-\mu_k}. \eqno (A.6)
$$
The leading vector is a polynomial satisfying the two equations
$$S^zv_J=Jv_J; \ \ S^+v_J=0. \eqno (A.7)
$$
These equations reduce to recurrence relations for the coefficients 
$a_{\mu_l,\mu_k}$ which have the unique solution
$$v_J=x_l^{2s_l-\mu}x_k^{2s_k-\mu}(y_lx_k-y_kx_l)^{\mu}. \eqno (A.8)
$$
Here $\mu =s_l+s_k-J$.

The remaining vectors of the representation $V_J$ can be obtained from 
the leading vector by means of the action of the lowering operator $S^-$. 
However, it should be noted that the operator $S$--commutes with the 
operator of multiplication by bracket in (A.8):
$$[S^-,y_lx_k-y_kx_l]=0. \eqno (A.9)
$$
>From this it is follows that all the vectors of the irreducible 
representation $V_J$ are divisible by $(y_lx_k-y_kx_l)^{\mu}$.

We have thus proved that if it is known that after addition of two spins 
$s_l$ and $s_k$ there is no projection onto states with full spin
$$s_l+s_k+1-\mu\le s_l+s_k,
$$
then all the vectors of the representation are divisible by 
$(y_lx_k-y_kx_l)^{\mu}$. We recall that $a^+_k=x_k$, $b^+_l=y_l$.

\vskip 0.5cm
{\bf Appendix B.}
\vskip 0.5cm

To compute the eigenvalues of the integral operator $\wh K_M$ for the 
Heisenberg $O(m)$--model on a one--dimensional chain we use properties of 
the Gegenbauer polynomials $C^{\lambda}_N(x)$. We recall [19] that
$$C^{\lambda}_N(x)=\sum_{l\ge 0}\frac{(-1)^l(\lambda )_{N-l}}{l!(N-2l)!}
(2x)^{N-2l}. \eqno (B.1)
$$
We assume below that $\lambda =m/2-1$. The relation between the 
Gegenbauer polynomials $C^{\lambda}_N(x)$ and harmonic polynomials 
$\Lambda^{A_N}({\bf n})$ defined by formulas (\ref{9.5}) and (\ref{12.3}) 
consists in the following equality:
$$\sum_{A_N}\Lambda^{A_N}_N({\bf n}_1)n_2^{A_N}=\frac{N!}{2^N(\lambda )_N}
C^{\lambda}_N({\bf n}_1\cdot{\bf n}_2). \eqno (B.2)
$$
Formula (B.2) can be verified directly. We note that
$$C^{\lambda}_N(1)=\frac{(2\lambda )_N}{N!}. \eqno (B.3)
$$
Hence,
$$\sum_{A_N}\Lambda^{A_N}_N({\bf n})n^{A_N}=\frac{N!}{2^N(\lambda )_N}
\left(\begin{array}{c}N+2\lambda -1\\ N\end{array}\right). \eqno (B.4)
$$
We now proceed to evaluate the integrals over the $(m-1)$--dimensional 
sphere of functions containing the Gegenbauer polynomials 
$C^{\lambda}_N({\bf n}_i\cdot{\bf n}_j)$. We first recall  (see [19]) the 
definition of spherical coordinates in ${\bf R}^m$ and the invariant 
measure on the sphere $S^{m-1}$. Cartesian and spherical coordinates are 
connected by the transformation
$$x_1=r\sin\theta_{m-1}\cdots\sin\theta_2\sin\theta_1,
$$
$$x_2=r\sin\theta_{m-1}\cdots\sin\theta_2\cos\theta_1,
$$
$$ .~.~.~.~.~.~.~.~.~.~.~.~.~.~. \eqno (B.5)
$$
$$x_{m-1}=r\sin\theta_{m-1}\cos\theta_{m-2},
$$
$$x_m=r\cos\theta_{m-1}.
$$
Here $0\le r\le\infty$, $0\le\theta_1<2\pi$, $0\le\theta_k<\pi$, $k\ne 1$.

The invariant measure on the sphere $S^{m-1}$ is defined by the formula
$$d\Omega=\frac{\Gamma (m/2)}{2\pi^{m/2}}\sin^{m-2}\theta_{m-1}\cdots
\sin\theta_2d\theta_1\cdots d\theta_{m-1}. \eqno (B.6)
$$
Using the transition to spherical coordinates, it is not hard to show that
\vskip 0.3cm

$\ds\int_{S^{m-1}}d\Omega_j\left(\frac{1-{\bf n}_i\cdot{\bf 
n}_j}{2}\right)^M ({\bf n}_i\cdot{\bf n}_j)^N$
\vskip 0.3cm
$$=\frac{\Gamma (2\lambda +1)\Gamma (M+\lambda +\frac{1}{2})\Gamma 
(N+\lambda +\frac{1}{2})}{[\Gamma (\lambda +\frac{1}{2})]^2\Gamma 
(M+N+2\lambda +1)}~_2F_1\left(\begin{array}{l}-N,M+\lambda +\frac{1}{2}\\
-N-\lambda +\frac{1}{2}\end{array}\Big\vert ~1\right) . \eqno (B.7)
$$
We recall that $\lambda =m/2-1$. Further, using (B.7) and (B.1), it is 
possible to evaluate the integral
\vskip 0.3cm

$\ds\int_{s^{m-1}}d\Omega\left(\frac{1-{\bf n}_i\cdot{\bf n}_j}{2}\right)^M
C_N^{\lambda}({\bf n}_i\cdot{\bf n}_k)$
\vskip 0.3cm

$$=\frac{\Gamma (2\lambda +1)\Gamma 
(M+\lambda +\frac{1}{2})(-M)_N}{\Gamma (\lambda +\frac{1}{2})\Gamma 
(M+N+2\lambda +1)}C_N^{\lambda}({\bf n}_i\cdot{\bf n}_k). \eqno (B.8)
$$
We now proceed to the proof of Theorem \ref{9.2} and formula 
(\ref{12.19}). To this end we consider the integral
$$\Theta ({\bf n}_i)=\int_{S^{m-1}}d\Omega\left(\frac{1-{\bf 
n}_i\cdot{\bf n}_j}{2}\right)^M\Lambda^{A_N}({\bf n}_j). \eqno (B.9)
$$
>From the Theorem \ref{t12.1}, 2) it follows that $\Theta ({\bf 
n}_i)=c\Lambda^{A_N}({\bf n}_i)$. In order to find the constant $c$, we 
consider the convolution (B.2) of the function $\Theta ({\bf n}_i)$ with 
$n_i^{A_N}$. By (B.8) we find
$$c=\frac{\Gamma (2\lambda +1)\Gamma (M+\lambda +\frac{1}{2})(-M)_N}
{\Gamma (\lambda +\frac{1}{2})\Gamma (M+N+2\lambda +1)}, \eqno (B.10)
$$
which coincides with the eigenvalues $Z_m(N,M)$ given by formula 
(\ref{12.20}). We formulate the result so obtained:
\vskip 0.3cm

$\ds\int_{S^{m-1}}d\Omega\left(\frac{1-{\bf 
n}_i\cdot{\bf n}_j}{2}\right)^M\Lambda^{A_N}({\bf n}_j)$
\vskip 0.3cm

$$=\frac{\Gamma (2\lambda +1)\Gamma (M+\lambda +\frac{1}{2})(-M)_N}
{\Gamma (\lambda +\frac{1}{2})\Gamma (M+N+2\lambda +1)}\Lambda^{A_N}({\bf 
n}_i). \eqno (B.11)
$$
We shall say a few words regarding the proof of Theorem \ref{t13.1}. The 
details will be published in a separate paper. From identity (B.11) and 
the inversion formula (\ref{12.13}) we obtain the equality

\vskip 0.3cm
$\ds\int_{S^{m-1}}d\Omega\left(\frac{1-{\bf 
n}_i\cdot{\bf n}_j}{2}\right)^Mn_j^{A_N}$
\vskip 0.3cm

$$=\frac{\Gamma (2\lambda +1)\Gamma (M+\lambda +\frac{1}{2})}{\Gamma 
(\lambda +\frac{1}{2})\Gamma (M+\lambda +1)}~\sum_{l=0}^{[N/2]}
Z_m^{(l)}(N,M)\sum_{A_{2l}\subset A_N}n^{A_N\setminus A_{2l}}\delta 
(A_{2l}), \eqno (B.12)
$$
where, by definition,
$$Z_m^{(l)}(N,M)=\frac{2^l(-M)_{N-2l}(M+\lambda +\frac{1}{2})}
{(M+2\lambda +1)_N}. \eqno (B.13)
$$
>From identity (B.12) it is easy to derive the equality

\vskip 0.3cm
$\ds\int_{S^{m-1}}d\Omega\left(\frac{1-{\bf 
n}_i\cdot{\bf n}_j}{2}\right)^M({\bf n}_j\cdot{\bf n}_{A_N})$
\vskip 0.3cm

$$=\frac{\Gamma (2\lambda +1)\Gamma (M+\lambda +\frac{1}{2})}{\Gamma 
(\lambda +\frac{1}{2})\Gamma (M+\lambda +1)}\sum_{A_{2l}\subset A_N}
Z_m^{(l)}(N,M)({\bf n}_i\cdot{\bf n}_{A_N\setminus A_{2l}})\delta 
({\bf n}_{A_{2l}}), \eqno (B.14)
$$
Here we have used the notation
$$({\bf n}_0\cdot{\bf n}_A)=\prod_{a\in A}({\bf n}_0\cdot{\bf n}_a),
$$
$$\delta ({\bf n}_{A_{2l}})=\frac{1}{l!}\sum_{A_{2l}\cup_{\alpha}\{ 
a_{\alpha},b_{\alpha}\}}\prod_{\alpha}({\bf n}_{a_{\alpha}}\cdot{\bf 
n}_{b_{\alpha}}) \eqno (B.15)
$$
(see definition (\ref{9.4})). We note the following consequence of 
formula (B.14):
$$\int_{S^{m-1}}d\Omega_0({\bf n}_0\cdot{\bf n}_{A_N})=2^{N/2}
\frac{\Gamma (2\lambda +1)\Gamma (M+\frac{N}{2}+\lambda +\frac{1}{2})}
{\Gamma (\lambda +\frac{1}{2})\Gamma (M+N+2\lambda +1)}~\delta ({\bf 
n}_{A_N}). \eqno (B.16)
$$
The next step is proving Theorem \ref{t13.1} consists in considering the 
integral operator $\wh K_{M_1.M_2}$: $C^{\infty}(S^{m-1})\to 
C^{\infty}(S^{m-1}\times S^{m-1})$:
$$(\wh K_{M_1,M_2}f)({\bf n}_1\cdot{\bf n}_2)=\int_{S^{m-1}}d\Omega
\left(\frac{1-{\bf n}_0\cdot{\bf n}_1}{2}\right)^{M_1}\left(\frac{1-{\bf 
n}_0\cdot{\bf n}_2}{2}\right)^{M_2}f({\bf n}_0). \eqno (B.17)
$$
It is clear that the case $p=2$ of Theorem \ref{t13.1} corresponds to the 
evaluation of the function $(\wh K_{M_1,M_2}{\bf 1})({\bf n}_1\cdot{\bf 
n}_2)$. This function can be most simply computed by using the expansion 
of the series $((1-x)/2)^M$ in the Gegenbauer polynomials 
$C^{\lambda}_N(x)$ for $\lambda =m/2-1$:
$$\left(\frac{1-x}{2}\right)^M=\frac{\Gamma (2\lambda +1)\Gamma 
(M+\lambda +\frac{1}{2})}{\Gamma (\lambda +\frac{1}{2})\Gamma (M+2\lambda 
+1)}\sum_{N\ge 0}\frac{(N+\lambda )(-M)_N}{\lambda (M+2\lambda +1)_N}
C_N^{\lambda}(x) \eqno (B.18)
$$
(see, for example, [19]). As a result, we obtain

\vskip 0.3cm
$\ds\int_{S^{m-1}}d\Omega
\left(\frac{1-{\bf n}_0\cdot{\bf n}_1}{2}\right)^{M_1}\left(\frac{1-{\bf 
n}_0\cdot{\bf n}_2}{2}\right)^{M_2}$
\vskip 0.3cm
$$=\frac{\Gamma (2\lambda +1)\Gamma (M_1+M_2+\lambda +\frac{1}{2})}
{\Gamma (\lambda +\frac{1}{2})\Gamma (M_1+M_2+2\lambda +1)}~_2F_1\left(
\begin{array}{l}-M_1,-M_2\\ -M_1-M_2-\lambda +\frac{1}{2}\end{array}
\Big\vert ~\frac{1-{\bf n}_1\cdot{\bf n}_2}{2}\right) . \eqno (B.19)
$$
It is clear that equality (B.19) corresponds to the case $p=2$ of the 
Theorem \ref{t13.1}. The next step consists in computing the function 
$\wh K_{M_1,M_2}({\bf n}_0\cdot{\bf n}_{A_N})$ for any set $A_N$. This 
computation is based on the expansion (B.18) and identity (B.14). We 
shall not present the formula for this function but rather immediately 
formulate a general result having Theorem \ref{t13.1} as a special case.

\vskip 0.2cm
{\bf Theorem B.} {\it Let ${\bf n}_0,{\bf n}_1,\ldots ,{\bf n}_p\in 
S^{m-1}$. Then}
\vskip 0.3cm

$\ds\int_{S^{m-1}}d\Omega_0\left(\frac{1-{\bf n}_0\cdot{\bf 
n}_1}{2}\right)^{M_1}\cdots\left(\frac{1-{\bf n}_0\cdot{\bf 
n}_p}{2}\right)^{M_p}({\bf n}_0\cdot{\bf n}_{A_N})$
\vskip 0.3cm

$$=\frac{\Gamma (2\lambda +1)\ds\prod_{j=1}^p\Gamma (M_j+1)(-1)^N}
{\Gamma (\lambda +\frac{1}{2})\Gamma (|M|+N+2\lambda +1)} \eqno (B.20)
$$
\begin{eqnarray}
&\times&\left\{\sum_{\begin{array}{c}k_1,\ldots ,k_p\\ N\equiv |k|~(\mod 2)
\end{array}}2^{1/2(N-|k|)}~\frac{\Gamma (|M|+\frac{1}{2}(N-|k|+\lambda 
+\frac{1}{2})}{\ds\prod_{j=1}^p\Gamma (M_j-k_j+1)}\right.\nonumber \\
&& \nonumber \\
&\times& F^{(p)}\left(\begin{array}{l}-M_1+k_1,\ldots ,-M_p+k_p\\ 
-|M|-\frac{1}{2}(N-|k|)_\lambda +\frac{1}{2}\end{array}~\Big\vert ~\left\{
\frac{1-{\bf n}_i\cdot{\bf n}_j}{2}\right\}\right)\nonumber \\
&\times& \left.\sum_{\begin{array}{c}\{ A_{k_j}\}\\ A_{k_j}\cap A_{k_i}=\O 
,\\ i\ne j\end{array}}\prod_{j=1}^p({\bf n}_j\cdot{\bf n}_{A_{k_j}})\delta
\left({\bf n}_{A_n}\setminus\ds\cup_{j=1}^pA_{k_j}\right)\right\} .\nonumber
\end{eqnarray}
Here we have used the notation (B.15), (\ref{13.8}) and 
$|M|=\ds\sum_{j=1}^pM_j$, $|k|=\ds\sum_{j=1}^pk_j$. From formula (B.20) 
we obtain an expression for the multipoint correlation functions of the 
classical Heisenberg $O(m)$--model on a complete graph (and hence on an 
arbitrary graph). The answer is a relation of generalized hypergeometric 
series which is too cumbersome and requires further study and 
simplification.

In conclusion we note that identity (B.18) gives an alternative means of 
computing the statistical sums $\langle\psi ~|~\psi\rangle_0$ and 
$\langle\psi ~|~\psi\rangle_{\rm reg}$ (see (\ref{12.23}) and 
(\ref{12.25})) of the classical Heisenberg $O(m)$--model on open and 
closed chains respectively.

\end{document}